\documentclass[conference]{IEEEtran}

\usepackage{epsfig}
\usepackage{tabularx}
\usepackage{colortbl}
\usepackage{boldline}
\usepackage{colortbl}
\usepackage{stackengine}
\usepackage{graphicx,times} 
\usepackage{color}
\usepackage{xspace}
\usepackage{listings}
\usepackage{fancyvrb}
\usepackage{verbatim}
\usepackage{booktabs}
\usepackage{colortbl}
\usepackage{relsize}
\usepackage{acronym}
\usepackage{paralist}
\usepackage{threeparttable}
\usepackage{flushend}
\usepackage{amsmath}
\usepackage{amsfonts}
\usepackage{amssymb}
\usepackage{booktabs}
\usepackage{subfigure}
\usepackage{ifthen}
\usepackage{url}
\usepackage{mathtools}
\usepackage{nccmath}
\usepackage{algorithm,algpseudocode}
\usepackage{amsfonts}
\usepackage[usenames,x11names]{xcolor}
\usepackage{tikz,pgfplots}
\usetikzlibrary{arrows}
\usepackage{pgfplotstable}
\usepgfplotslibrary{units,groupplots}
\usepackage{wrapfig}
\usepackage{varwidth,xcolor}
\usepackage{setspace}
\usepackage{capt-of}
\usepackage{multirow}
\usepackage{hhline}
\usepackage{fancyhdr}

\newcommand{\highlightnewtext}{0}
\newcommand{\keepoldtext}{0}

\ifnum\highlightnewtext=1
	\newcommand{\revised}[1]{{\leavevmode\color{blue}{#1}}}
\else
	\newcommand{\revised}[1]{{\leavevmode\color{black}{#1}}}
\fi

\ifnum\keepoldtext=1
\newcommand{\original}[1]{{\leavevmode\color{orange}{#1}}}
\else
\newcommand{\original}[1]{}
\fi

\makeatletter
\let\myorg@bibitem\bibitem
\def\bibitem#1#2\par{%
	\@ifundefined{bibitem@#1}{%
		\myorg@bibitem{#1}#2\par
	}{%
		\begingroup
		\color{\csname bibitem@#1\endcsname}%
		\myorg@bibitem{#1}#2\par
		\endgroup
	}%
}
\ifnum\highlightnewtext=1
	\newcommand*{\bibitem@silo}{blue}
	\newcommand*{\bibitem@headerspace}{blue}
	\newcommand*{\bibitem@huang}{blue}
	\newcommand*{\bibitem@nfvipop}{blue}
\fi
\makeatother

\newcommand{\mytinysize}{\fontsize{6}{7}\selectfont}
\pgfplotsset{
	compat = newest,
	tick label style={font=\sffamily\scriptsize},
	label style={font=\sffamily\scriptsize},
	legend style={font=\sffamily\mytinysize\raggedleft},
	legend cell align=left,
	grid style={dotted,gray}
}

\definecolor{mygray}{RGB}{220,220,220}
\definecolor{diffcolor}{RGB}{0,0,0}

\makeatletter
\algrenewcommand\ALG@beginalgorithmic{\footnotesize} 
\makeatother

\newcolumntype{L}[1]{>{\raggedright\let\newline\\\arraybackslash\hspace{0pt}}m{#1}}
\newcolumntype{C}[1]{>{\centering\let\newline\\\arraybackslash\hspace{0pt}}m{#1}}
\newcolumntype{R}[1]{>{\raggedleft\let\newline\\\arraybackslash\hspace{0pt}}m{#1}}

\graphicspath{{artworks/}{graphs/}}

\begin{document}

\title{Dynamic and Application-Aware Provisioning of Chained Virtual Security Network Functions}

\author{
	Roberto Doriguzzi-Corin$^\alpha$,
	Sandra Scott-Hayward$^\beta$,
	Domenico Siracusa$^\alpha$,
	Marco Savi$^\alpha$,
	Elio Salvadori$^\alpha$\\
	
	\small{$^\alpha$CREATE-NET, Fondazione Bruno Kessler - Italy}\\
	\small{$^\beta$CSIT, Queen's University Belfast - Northern Ireland}
	
}

\maketitle

\thispagestyle{fancy}
\renewcommand{\headrulewidth}{0pt}
\chead{\scriptsize This is the author's version of an article that has been published in IEEE Transactions on Network and Service Management. Changes were made to\\this version by the publisher prior to publication. The final version of record is available at {\color{blue}https://doi.org/10.1109/TNSM.2019.2941128}. \\ The source code associated with this project is available at {\color{blue}{https://github.com/doriguzzi/pess-security}}.}
\cfoot{\scriptsize Copyright (c) 2019 IEEE. Personal use is permitted. For any other purposes, permission must be obtained from the IEEE by emailing pubs-permissions@ieee.org.}

\begin{abstract}    
	A promising area of application for Network Function Virtualization is in network security, where chains of Virtual Security Network Functions (VSNFs), i.e., security-specific virtual functions such as firewalls or Intrusion Prevention Systems, can be dynamically created and configured to inspect, filter or monitor the network traffic. However, the traffic handled by VSNFs could be sensitive to specific network requirements, such as minimum bandwidth or maximum end-to-end latency. Therefore, the decision on which VSNFs should apply for a given application, where to place them and how to connect them, should take such requirements into consideration. Otherwise, security services could affect the quality of service experienced by customers.
	
	{\color{diffcolor} In this paper we propose PESS (Progressive Embedding of Security Services), a solution to efficiently deploy chains of virtualised security functions based on the security requirements of individual applications and operators' policies, while optimizing resource utilization.}
	We provide the PESS mathematical model and heuristic solution. 
	
	{\color{diffcolor}Simulation results show that, compared to state-of-the-art application-agnostic VSNF provisioning models, PESS reduces computational resource utilization by up to 50\%, in different network scenarios.}
	This result ultimately leads to a higher number of provisioned security services and to up to a 40\% reduction in end-to-end latency of application traffic.
\end{abstract}

\begin{IEEEkeywords}
\revised{\acl{nfv}, \acl{nsc}, Progressive Embedding, Application-Aware Network Security.}
\end{IEEEkeywords}


\acrodef{acl}[ACL]{Access Control List}
\acrodef{api}[API]{Application Programming Interface}
\acrodef{cctv}[CCTV]{Closed Circuit Television}
\acrodef{cpe}[CPE]{Customer Premise Equipment}
\acrodef{dlp}[DLP]{Data Loss/Leakage Prevention}
\acrodef{dpi}[DPI]{Deep Packet Inspection}
\acrodef{ddos}[DDoS]{Distributed Denial of Service}
\acrodef{foss}[FOSS]{Free and Open-Source Software}
\acrodef{ha}[HA]{Hardware Appliance}
\acrodef{ids}[IDS]{Intrusion Detection System}
\acrodef{ilp}[ILP]{Integer Linear Programming}
\acrodef{isp}[ISP]{Internet Service Provider}
\acrodef{ips}[IPS]{Intrusion Prevention System}
\acrodef{mano}[NFV MANO]{NFV Management and Orchestration}
\acrodef{mips}[MIPS]{Millions of Instructions Per Second}
\acrodef{ml}[ML]{Machine Learning}
\acrodef{nat}[NAT]{Network Address Translation}
\acrodef{nf}[NF]{Network Function}
\acrodef{nfv}[NFV]{Network Function Virtualization}
\acrodef{nsc}[NSC]{Network Service Chaining}
\acrodef{of}[OF]{OpenFlow}
\acrodef{pess}[PESS]{Progressive Embedding of Security Services}
\acrodef{pop}[PoP]{Point of Presence}
\acrodef{ps}[PS]{Port Scanner}
\acrodef{qoe}[QoE]{Quality of Experience}
\acrodef{qos}[QoS]{Quality of Service}
\acrodef{sdn}[SDN]{Software-Defined Networking}
\acrodef{sla}[SLA]{Service Level Agreement}
\acrodef{snf}[SNF]{Security Network Function}
\acrodef{tc}[TC]{Traffic Classifier}
\acrodef{tor}[ToR]{Top of Rack}
\acrodef{tsp}[TSP]{Telecommunication Service Provider}
\acrodef{vm}[VM]{Virtual Machine}
\acrodef{vne}[VNE]{Virtual Network Embedding}
\acrodef{vnep}[VNEP]{Virtual Network Embedding Problem}
\acrodef{vnf}[VNF]{Virtual Network Function}
\acrodef{vsnf}[VSNF]{Virtual Security Network Function}
\acrodef{vpn}[VPN]{Virtual Private Network}
\acrodef{wan}[WAN]{Wide Area Network}
\acrodef{waf}[WAF]{Web Application Firewall}

\section{Introduction}\label{sec:introduction}
Network security implemented by \acp{tsp} has traditionally been based on the deployment of specialized, closed, proprietary \acp{ha}. Such \acp{ha} are inflexible in terms of functionalities and placement in the network, which means that even slight changes in the security requirements generally necessitate manually intensive and time-consuming re-configuration tasks, the replacement of existing \acp{ha} or the deployment of additional \acp{ha}. 

The \ac{nfv}~\cite{7243304} initiative  has been proposed as a possible solution to address the operational challenges and high costs of managing proprietary \acp{ha}. The main idea behind \ac{nfv} is to transform network functions (e.g. firewalls, intrusion detection systems etc.) based on proprietary \acp{ha}, into software components (called \acp{vnf}) that can be deployed and executed in virtual machines on commodity, high-performance servers. By decoupling software from hardware, this approach allows any (security) network function to be deployed in any server connected to the network through an automated and logically centralized management system.

The centralized management system, called \ac{mano}, controls the whole life-cycle of each \ac{vnf}. In addition, the \ac{mano} can  dynamically provision complex network services in the form of sequences (often called chains) of \acp{vnf}.
Indeed, \ac{nsc} is a technique for selecting subsets of the network traffic and forcing them to traverse various \ac{vnf}s in sequence. For example, a firewall followed by an \ac{ips}, then a \ac{nat} service and so on. \ac{nsc} and \ac{nfv} enable flexible, dynamic service chain modifications to meet the real time network demands.

A promising area of application for \ac{nsc} and \ac{nfv} is in network security, where chains of \acfp{vsnf}, i.e., security-specific \acp{vnf} such as a firewall or an \ac{ips}, can be dynamically created and configured to inspect, filter or monitor the network traffic. The flexibility of the \ac{nsc} and \ac{nfv} paradigms brings many benefits, among others: (i) highly customizable security services based on the needs of the end-users, (ii) fast reaction to new security threats or variations of known attacks, and (iii) low Operating Expenditure (OpEx) and Capital Expenditure (CapEx) for the operator. On the other hand, compared to specialized \acp{ha}, \acp{vsnf} may have a significant impact on the performance of the network and on the \ac{qos} level experienced by the users. 
The virtualization overhead, the utilization level of the servers and the techniques adopted to implement the \acp{vsnf} are the most significant contributors to the QoS degradation.

\begin{table*}[h!]
	\centering
	\scriptsize
	\caption{Security and \ac{qos} requirements of applications.}
	\label{tab:tspapps}
	\renewcommand{\arraystretch}{1.1}
	\begin{threeparttable}
		\begin{tabular}{V{2}>{\bfseries}c|c|c|c|c V{2}}
			\hlineB{2}
			\rowcolor{mygray}
			\textbf{Application class} & \textbf{Description} & \textbf{Related threats}
			& \textbf{Relevant \ac{vsnf} \tnote{1}} & \textbf{\ac{qos} requirements}\\ \hlineB{2}
			CCTV systems  & \begin{tabular}{@{}c@{}}Closed Circuit TV\\for video surveillance\\accessible remotely \end{tabular}  &   \begin{tabular}{@{}c@{}}Port scanning, DDoS\\ password cracking \end{tabular}   & \begin{tabular}{@{}c@{}}Firewall, DPI,\\ IDS, IPS \end{tabular}     &  \begin{tabular}{@{}c@{}}Bandwidth: 10Mbps\\ (5 cameras, 720p, 15fps,\\ H.264, medium quality)\\ Latency: 200ms (PTZ\tnote{2} \\two-way latency~\cite{5384976})\end{tabular}  \\ \hline
			Email  & Electronic mail  &   \begin{tabular}{@{}c@{}}Malware, spam, phishing, data exfiltration  \end{tabular}   & \begin{tabular}{@{}c@{}}DPI, Antispam, IDS, DLP\end{tabular}   &   --   \\ \hline
			Instant messaging  & \begin{tabular}{@{}c@{}}Real-time text-based Internet chat\end{tabular}      & \begin{tabular}{@{}c@{}}Malware, DDoS, phishing (out-of-band)\end{tabular}              & \begin{tabular}{@{}c@{}}DPI, Antispam, IDS, IPS\end{tabular}     &  --   \\ \hline
			\begin{tabular}{@{}c@{}}Media streaming\end{tabular}  & \begin{tabular}{@{}c@{}}Audio/video content\\ accessed over the Internet\end{tabular}  &   \begin{tabular}{@{}c@{}}Inappropriate content  \end{tabular}   & \begin{tabular}{@{}c@{}}Parental control\end{tabular}   &   \begin{tabular}{@{}c@{}} Bandwidth\tnote{3}: 5Mbps (HD)\\25Mbps (UHD)  \end{tabular}   \\ \hline
			\begin{tabular}{@{}c@{}}Remote storage\end{tabular}  & \begin{tabular}{@{}c@{}}File transfer over the network\end{tabular}  &   \begin{tabular}{@{}c@{}}Data exfiltration  \end{tabular}   & \begin{tabular}{@{}c@{}}VPN, Data Encryption\end{tabular}   &   \begin{tabular}{@{}c@{}} Bandwidth  \end{tabular}   \\ \hline
			\begin{tabular}{@{}c@{}}Network services\\ (DNS, VoD, file\\sharing, WWW, SSH)\end{tabular}  & \begin{tabular}{@{}c@{}}Server application\\ accessed by remote\\ client applications\end{tabular}  &   \begin{tabular}{@{}c@{}}DDoS, SQL injection,\\ remote code execution  \end{tabular}   & \begin{tabular}{@{}c@{}}Firewall, IDS,\\ WAF, Honeypot\end{tabular}     &  --  \\ \hline
			Online gaming  & \begin{tabular}{@{}c@{}}Video games played\\ over the Internet\end{tabular}             &   \begin{tabular}{@{}c@{}}Online game cheating (out-of-band attacks)\\ DDoS (in-band attacks)\end{tabular}     & \begin{tabular}{@{}c@{}}DPI, Antispam,\\ IDS, IPS\end{tabular}    &  \begin{tabular}{@{}c@{}}Latency: 100ms\\ (first-person games~\cite{Claypool:2006:LPA:1167838.1167860})\end{tabular}   \\  \hline
			Peer-to-peer  & \begin{tabular}{@{}c@{}}File sharing over peer-to-peer networks\end{tabular}               &   DDoS, malware  & DPI, IDS, IPS   &   --  \\ \hline
			\begin{tabular}{@{}c@{}}Video conferencing\end{tabular}  & \begin{tabular}{@{}c@{}}Real-time audio/video over the Internet\end{tabular}  &   \begin{tabular}{@{}c@{}}DDoS  \end{tabular}   & \begin{tabular}{@{}c@{}}Firewall, IPS\end{tabular}   &   \begin{tabular}{@{}c@{}} Latency: 150ms~\cite{Chen:2004:QRN:1234242.1234243}  \end{tabular}   \\ \hline
			Web browsing  & \begin{tabular}{@{}c@{}}Applications for browsing \\the  World Wide Web\end{tabular}         &   \begin{tabular}{@{}c@{}} Cross-site scripting, phishing,\\ malware, inappropriate content \end{tabular}     & \begin{tabular}{@{}c@{}} DPI, WAF,\\ Parental control\end{tabular}    &   \begin{tabular}{@{}c@{}} Latency: 400ms~\cite{Chen:2004:QRN:1234242.1234243} \end{tabular}  \\ \hlineB{2}
		\end{tabular}
		\begin{tablenotes}
			\item $^1$Acronyms of \acp{vsnf}: \acf{dpi}, \acf{ids}, \acf{ips}, \acf{dlp}, \acf{vpn}, \acf{waf}. $\quad^2$PTZ: Pan, Tilt, Zoom. $\quad^3$Netflix Internet connection speed recommendations.
		\end{tablenotes}
	\end{threeparttable}
\end{table*}

We argue that, for a wide adoption of \ac{nsc} and \ac{nfv} technologies in network security, the provisioning strategies should take into account not only the security requirements, but also specific \ac{qos} needs of applications (a list of common classes of applications is provided in Table \ref{tab:tspapps} with their corresponding security and QoS requirements). Omitting the latter may lead, for instance, to a model that blindly forces all the user traffic to traverse the whole chain of \acp{vsnf}. As a result, computationally demanding \acp{vsnf} such as \acp{ips} may cause a noticeable performance degradation to latency-sensitive applications (e.g. online games~\cite{Claypool:2010:LKP:1730836.1730863}) or bandwidth sensitive applications (e.g. video streaming). On the other hand, beyond the overall resource consumption and \ac{qos} requirements, the model must also take into account the specific security best practices and policies. Omitting such aspects may result in the inappropriate placement of a firewall in the middle of the network, thus allowing unauthorized traffic to reach hosts that should be protected.

In this paper, we present \acs{pess} (\aclu{pess}), a novel approach to provision security services by composing chains of VSNFs according to the specific QoS needs of user applications and the security policies defined by the \ac{tsp}. \ac{tsp}'s security policies (given as an input to \ac{pess}) include: the kind of \acp{vsnf} (e.g., firewall, \ac{ips}, etc.) that should be deployed for a specific class of applications, their order (e.g., firewall first, then \ac{ips}, etc.), and more (e.g., a parental control should be installed close to the user's premises). 

\ac{pess} defines an \ac{ilp} formulation and a heuristic algorithm to tackle the provisioning problem in dynamic network scenarios, where the service requests are not known in advance. In contrast, advance knowledge of service requests is assumed by the majority of related works. 
Although the \ac{pess} formulation and implementation presented in this paper focus on security-specific services, the proposed approach is also suitable for more complex scenarios, where heterogeneous network services provided by means of generic \acp{vnf} coexist (e.g., security, video broadcasting, content caching, etc.). 

To the best of our knowledge, this work is the first attempt to tackle the challenges of progressively provisioning application-aware security services onto operational infrastructures, in which the \ac{qos} performance of existing services may be compromised by adding new services. 

A preliminary \ac{ilp} formulation of our application-aware approach has been introduced in \cite{short-paper}. This work extends \cite{short-paper} with the following contributions:
\begin{itemize}
	{\color{diffcolor}
	\item A taxonomy of the most relevant \acp{vsnf} along with a list of popular implementations available in the open-source community and in today's market. 
	}
	\item A mathematical formulation for the progressive provisioning of security services (the \ac{pess} \ac{ilp} model). With respect to the \ac{ilp} model presented in \cite{short-paper}, this work formulates the estimation of the processing delay based on residual computing resources of the physical nodes, and it formalizes the impact on the end-to-end latency of operational services when allocating computing resources for new requests.
	\item A heuristic algorithm, called \ac{pess} heuristic, to obtain near-optimal solutions of the embedding problem in an acceptable time frame (in the order of a few milliseconds even in large network scenarios).
	\item An evaluation of the heuristic's performance in terms of quality of the solutions (deviation from optimality) and scalability performed on real-world and randomly generated topologies.
	\item Comparison of the proposed application-aware approach against the baseline approach, in which security services are provided without taking into account the specific requirements of applications.
\end{itemize}

{\color{diffcolor} 	
Our simulations prove that the \ac{pess} solution can deploy more security services over the same infrastructure compared to an application-agnostic approach (the baseline), while still respecting the security policies and best practices defined by the \ac{tsp}.
With PESS, each traffic flow generated by each application can be served by the strict subset of \acp{vsnf} that are necessary to ensure its security, which means that no flow is burdened with any unnecessary security function that could affect its smooth execution, as would be the case with an application-agnostic approach. 
Of course, the capability of the \acp{vsnf} to properly react to any security attack depends on the specific implementation of the \ac{vsnf} itself and is beyond the scope of this work.}

The remainder of this paper is structured as follows: Section \ref{sec:sota} reviews and discusses the related work. {\color{diffcolor} Section \ref{sec:background} gives the relevant background information. Section \ref{sec:motivation} provides the motivation behind this work.} 
Section \ref{sec:model} details the mathematical formulation of the \ac{ilp} model, while Section \ref{sec:implementation} describes the heuristic algorithm that we implemented to solve the problem. In Section \ref{sec:evaluation}, the heuristic algorithm is evaluated on real-world and random topologies.  Finally, the conclusions are provided in Section \ref{sec:conclusions}.

\section{Related work}\label{sec:sota}
With the recent ``softwarisation'' of network resources, a plethora of research initiatives has emerged in the last few years to address the problem of the optimal placement of chained \acp{vnf}. Most of these tackle the problem by using linear programming techniques and by proposing heuristic algorithms to cope with large scale problems. In this section, we classify and review the most relevant works for our studies.

\subsection{QoS-driven \ac{vnf} Placement}
\ac{qos}-driven approaches primarily focus on the \ac{qos} requirements of specific services without considering network security aspects. In this regard, the proposed mathematical models include bandwidth and latency constraints (similar to Constraints (\ref{eq:const_8}) and (\ref{eq:const_latency}) presented in Section \ref{sec:model}) or define objective functions that require minimization of the total bandwidth and latency of created chains.

The \ac{ilp} model in \cite{7469866} considers computing and bandwidth constraints to minimize the costs related to (i) \ac{vnf} deployment, (ii) energy consumption of the servers, and (iii) forwarding traffic. The end-to-end delay requirement is formulated as a penalty in the objective function. However, the computation of the end-to-end delay only considers link propagation delays without including the processing delay introduced at each \ac{vnf}. In~\cite{6968961}, the placement problem is formulated as a Mixed Integer Quadratically Constrained Problem with respect to bandwidth, number of used nodes and latency. The processing delay at each \ac{vnf} is also not considered in this work. 
The study in \cite{qos-driven} proposes an \ac{ilp} formulation and a heuristic algorithm for the \ac{vnf} placement problem focusing on \ac{qos} parameters such as end-to-end delay and \ac{nsc} availability. The \ac{ilp} model formulation presented in the paper does not discuss how the processing delay introduced by the \acp{vnf} is computed. This limitation is reflected in the assumptions made for the evaluation, where the processing delay is considered independent from the \ac{vnf} type/implementation and from the computing capacity of the physical node where \acp{vnf} are placed. 
{\color{diffcolor}
In \cite{8480442}, Tajiki et al. present a resource allocation architecture for softwarized networks. The proposed architecture includes two resource allocation modules whose goal is configuring the network while satisfying \ac{qos} constraints and optimizing the energy consumption and the number of flow entries in the network. Although the authors tackle the problem of progressively allocating resources for newly arrived flows, neither the \ac{ilp} formulation nor the heuristic algorithm consider the effects of the resource allocation on servers whose computing capacity is close to the limit. As discussed in Sections \ref{sec:constraints} and \ref{sec:implementation}, this may lead to a degradation of the \ac{qos} of existing services in terms of higher end-to-end latency.
}

\subsection{Placement of \acp{vnf}/\acp{vsnf}}\label{sec:relsecb}
In addition to the research work on QoS-driven \acp{vnf} placement, there are a number of works that specifically consider the placement of \acp{vsnf}. 


The method proposed in~\cite{Park:2017:DDP:3040992.3041005} is based on light-weight, protocol-specific intrusion detection \acp{vnf}. The system dynamically invokes a chain of these \acp{ids} according to the traffic characteristics. The placement of the chains is based on a user-defined or common shortest-path algorithm such as Dijkstra, without consideration of the application \ac{qos} requirements or available network/computing resources. 

In~\cite{7899497}, the authors argue that reactive mechanisms used by cloud providers to deploy \acp{vsnf} do not ensure an optimal resource allocation. To address this, the authors propose a novel resource allocation scheme, which estimates the behaviour of the traffic load by monitoring the history of the current \acp{vsnf}, and pro-actively provisions new instances of those \acp{vsnf} as a countermeasure to any incoming resource pressure. The proposed algorithm does not tackle the problem of \ac{vsnf} chaining. Instead, it focuses on the optimal placement of new instances of \acp{vsnf}, which are part of existing chains. It also assumes infinite network and computing resources.

In~\cite{8530989}, Dermici et al. tackle the \acp{vsnf} placement problem by proposing an \ac{ilp} formulation whose objective is the minimization of the energy consumption of servers. This solution does not consider any security nor \ac{qos} constraints. The aim of the \ac{nsc} embedding model presented in \cite{8254344} is to minimize the end-to-end latency of cross-domain chains of \acp{vsnf}. The main limitation of the proposed \ac{ilp} formulation is that it only considers link propagation delays, while ignoring the processing delay introduced at each \ac{vsnf}.

\begin{table*}[!t]
	\centering
	\scriptsize
	\caption{Taxonomy of security VNFs.}
	\label{tab:vnf_taxonomy}
	\renewcommand{\arraystretch}{1.1}
	\begin{threeparttable}
		\begin{tabular}{V{2}>{\bfseries}c|c|c|c V{2}}
			\hlineB{2}
			\rowcolor{mygray}
			\textbf{VNF} & \textbf{Description} & \textbf{Use in security} & \textbf{Implementations}  \\ \hlineB{2}
			Antispam  & Email filtering             &   \begin{tabular}{@{}c@{}}Malware detection, Phishing prevention\end{tabular}     & \begin{tabular}{@{}c@{}}SpamAssassin, rspamd, ASSP, Juniper vSRX\end{tabular} \\ \hline
			Antivirus  & Email/Web scanning/Endpoint security     &   Virus/Trojan/Malware detection    & \begin{tabular}{@{}c@{}}ClamAV, ClamWin, Juniper vSRX\end{tabular} \\ \hline
			DLP  & Data Loss/Leakage Prevention             &   Data exfiltration detection    & myDLP, OpenDLP
			\\ \hline
			DPI  & Payload analysis      & \begin{tabular}{@{}c@{}}Spam Filtering, Intrusion detection, DDoS\\detection, Malware detection, Security Analytics\end{tabular}              & \begin{tabular}{@{}c@{}}OpenDPI, nDPI, L7-filter, Libprotoident,\\PACE, NBAR, Cisco ASAv\end{tabular}          \\ \hline
			Honeypot  & \begin{tabular}{@{}c@{}}Traffic redirection and inspection\end{tabular}             &   \begin{tabular}{@{}c@{}}Spam filtering, Malware detection,\\ SQL database protection, Security Analytics \end{tabular}    & \begin{tabular}{@{}c@{}}HoneyD, SpamD, Kippo,\\Kojoney, Dionaea, Glastopf \end{tabular}        \\ \hline
			IDS  & \begin{tabular}{@{}c@{}}Traffic inspection \\ (header and payload)\end{tabular}            &   \begin{tabular}{@{}c@{}}Intrusion detection, Malware detection,\\DDoS detection, Security Analytics \end{tabular}   & \begin{tabular}{@{}c@{}}Snort, Bro, Suricata, ACARM-ng, AIDE,\\OSSEC, Samhain, Cuckoo, Cisco ASAv\end{tabular}         \\ \hline
			IPS  & \begin{tabular}{@{}c@{}}Traffic filtering based on\\ header and payload\end{tabular}             &   \begin{tabular}{@{}c@{}}Intrusion prevention,\\ DDoS prevention \end{tabular}    & \begin{tabular}{@{}c@{}}Snort, Suricata, ACARM-ng,\\Fail2Ban, Juniper vSRX \end{tabular}        \\ \hline
			NAT\tnote{1}  & IP address mapping               &   Intrusion prevention  & Netfilter, IPFilter, PF        \\ \hline
			Packet Filter Firewall  & \begin{tabular}{@{}c@{}}Header-based packet filtering\end{tabular}         &    Intrusion prevention      & \begin{tabular}{@{}c@{}}Netfilter, nftables, NuFW, IPFilter, ipfw, PF,\\Juniper vSRX, VMWare vShield, Fortigate FW\end{tabular}         \\ \hline
			\begin{tabular}{@{}c@{}}Parental Control\end{tabular}  & \begin{tabular}{@{}c@{}}Media content filtering\end{tabular}             &   \begin{tabular}{@{}c@{}}Blocking access to\\ inappropriate content\end{tabular}      & \begin{tabular}{@{}c@{}}OpenDNS, SquidGuard,\\ DansGuardian, pfsense\end{tabular}         \\ \hline
			VPN Gateway &  \begin{tabular}{@{}c@{}}Site-to-site VPN connection\\over unsecured networks\end{tabular}  & Data Tunneling/Encryption &   \begin{tabular}{@{}c@{}}OpenVPN, strongSwan, Juniper vSRX,\\Cisco ASAv, Fortigate VPN\end{tabular}    \\ \hline
			WAF  & \begin{tabular}{@{}c@{}}HTTP traffic monitoring, filtering, logging\end{tabular}             &   \begin{tabular}{@{}c@{}}Prevention of SQL injection, cross-site scripting\end{tabular}      & ModSecurity         \\ \hlineB{2}
		\end{tabular}
		\begin{tablenotes}
			\item[1] NAT is not a security function but inherently provides packet filtering similar to a firewall.
		\end{tablenotes}
	\end{threeparttable}	
\end{table*}

\subsection{Security-driven VSNF Placement}
Although the literature reviewed in Section \ref{sec:relsecb} addresses the placement of \acp{vsnf}, few solutions have been proposed with a focus on the network security requirements of the \ac{vsnf} placement. In \cite{7592416}, the authors propose a model for the placement of \acp{vsnf} that takes into account security deployment constraints. Such constraints are necessary to avoid incorrect deployment of security functions such as placing an \ac{ids} on an encrypted channel. The authors propose an \ac{ilp} formulation of the problem and validate their model by measuring the execution time in four different scenarios and by comparing the model with other heuristics in terms of placement cost. However, the proposed optimization algorithm is always computed for all flows in the network. Therefore, it does not scale well. The authors mitigate the problem by partitioning the network into independent blocks. Nevertheless, the partitioning scheme is limited to fat-tree topologies. Furthermore, the end-to-end latency is not considered among the constraints of the proposed model, which limits its application space.
{\color{diffcolor}The authors of \cite{8466784} propose an \ac{ilp} formulation and a heuristic algorithm for efficiently composing chains of virtual security functions. The \ac{ilp} formulation includes a single security-related constraint to ensure that the security level of each deployed \ac{vsnf} instance is higher than the security level required by the service request. However, this work does not take into account basic security aspects, such as order and operational mode (stateful/stateless) of the chained \acp{vsnf}. Moreover, the proposed formulation does not consider the mutual interference between security services caused by the concurrent access to the (finite) computing resources available in the infrastructure. The latter aspect is particularly relevant in a \ac{tsp} scenario \revised{(see also part II of \cite{huang})}, where the security services are provisioned in a dynamic manner based on the incoming customers' requests.}

\section{Background}\label{sec:background}
This work is underpinned by two emerging network technologies; \acf{nfv} and \revised{\ac{sdn}} and their integration to provision network security solutions.


\subsection{\acl{nfv}}\label{sec:nfv}
Today's network functions such as firewalling, \ac{dpi}, \acp{ids}, etc. are provided by specialized proprietary hardware appliances (also called \textit{middleboxes}) strategically deployed in the network.
The \ac{nfv} paradigm separates the network functions from the underlying hardware by moving the functions from specialized devices to off-the-shelf commodity equipment such as industry standard servers or high-performance network devices. Therefore, network services can be decomposed into multiple \acp{vnf} running on physical or virtual machines, which could be located in data centers, network nodes or at the end-user premises. 

In contrast to middleboxes, the configuration of which requires intensive and time-consuming manual intervention, \ac{nfv} allows an automated and flexible deployment of network functions on any NFV-enabled device. The lifecycle management of \acp{vnf} and hardware resources is achieved through a centralized software component called the Orchestrator.

\subsection{\revised{\acl{sdn}}}
\ac{sdn} is often referred to as a paradigm for network environments where the control plane is physically separated from the data plane and a logically centralized control plane controls several devices. This differs from traditional networks in which nodes are autonomous systems unaware of the overall state of the network. In \ac{sdn} deployments, the nodes are remotely controlled via standard protocols (e.g. OpenFlow~\cite{openflow}) by a logically centralized intelligent module called the \textit{\ac{sdn} controller}, which bases routing decisions on a global (domain) view of the network.\\
The controller is a software component which runs on commodity hardware appliances and provides an open \ac{api} to program the network for configuration, monitoring and troubleshooting purposes. Such programmability enables automated and dynamic network configurability and fine-grained control of the traffic based on the values of the packets' header fields (e.g., source/destination IP/MAC addresses, VLAN tags, TCP/UDP ports, etc.). 

\subsection{Service Function Chaining}
Service Function Chaining (also known as Network Service Chaining) is a technique for selecting and steering data traffic flows through network services. The network services can be traffic management applications such as load balancing, or security applications such as those detailed in Section~\ref{sec:vsnf}. Service function chaining combines the capabilities of \ac{sdn} and \ac{nfv} to connect a distributed set of \acp{vnf}.

\subsection{A Taxonomy of security \acp{vnf}}\label{sec:vsnf}
As introduced in Section~\ref{sec:nfv}, a \ac{vnf} is a software implementation of a network function which is deployed on a virtual resource such as a Virtual Machine. 
Table \ref{tab:vnf_taxonomy} provides a list of the most common security functions. Traditionally, the majority of these functions would have been implemented on dedicated hardware (middleboxes) to process the network traffic along the data path. Today, these functions are deployed as \acp{vnf}. Table \ref{tab:vnf_taxonomy} includes a short description of each \ac{vnf} and some of the publicly available open-source implementations or commercial products.

\section{Motivation}\label{sec:motivation}

We motivate our work by describing two use case scenarios, namely web browsing and online gaming, where the \ac{tsp} exploits the \ac{nsc} and \ac{nfv} technologies to provide security services tailored to specific users' application requirements.

\textbf{Web browsing.} Parental control is applied to Web traffic to block unwanted media and social-media content, while an \ac{ids} might be used to intercept malicious software (malware). Stateful \acp{vsnf} provide security functionality by tracking the state of network connections (e.g. Layer 4 firewall, \ac{nat}). In this case, the same \ac{vsnf} instance must be traversed by all traffic flows of a network conversation in order to maintain the correct state of the connection. More flexible provisioning schemes can be adopted for stateless \acp{vsnf}, where multiple instances of the same \ac{vsnf} might be deployed on different servers for load balancing. This example also illustrates the security best-practice that unwanted traffic should be blocked as soon as it enters the network by placing firewalls and \ac{ids}/\ac{ips} close to the border of the \ac{tsp} domain. Another generally accepted practice, is to place firewalls before \ac{ids}/\ac{ips} (from the point of view of incoming traffic). Firewalls are generally designed to drop unauthorized traffic very quickly, thus reducing the burden on \ac{ids}/\ac{ips}, which are more computationally expensive.

\textbf{Online gaming.} An \ac{ids} might also be used to detect possible threats due to the misuse of chat tools integrated within the gaming software (e.g., phishing~\cite{Gianvecchio:2011:HBI:2109150.2109174}, social engineering~\cite{Yan:2005:SCC:1103599.1103606}, etc.). As the communication between the client and the server relies on timely delivery of packets, \ac{ids} operations are not executed on the \textit{in-game} traffic. In this case the security is enforced by a faster \ac{vsnf} such as a Firewall, which checks the packet headers without any deep-payload analysis. It should be noted that web traffic and chat conversations are often encrypted by TLS/SSL cryptographic protocols. Although encryption preserves the confidentiality of the traffic, it also prevents \ac{ids}-based \acp{vsnf} such as Parental Control and \ac{ids} from inspecting the packets, thus allowing an attacker to obfuscate malicious data in encrypted payloads. However, the \ac{tsp} could overcome this limitation either using a Transparent Proxy \ac{vsnf} or by exploiting recent advances in network security~\cite{Sherry:2015:BDP:2785956.2787502,Canard:2017:BMP:3052973.3053013}.\\


These are just two examples of how the security service can be tailored to the user's application requirements by appropriate selection and placement of \acp{vsnf}. A list of common classes of applications supported in the \ac{tsp} use-case is provided in Table \ref{tab:tspapps} with their corresponding security and QoS requirements and relevant \acp{vsnf}. 
{\color{diffcolor}
One of them, the remotely accessible CCTV system, will be used in the rest of this paper as a running example to illustrate various aspects of our work.

According to the motivations provided above, we can summarize the rationale behind the \ac{pess} approach as follows: (i) a user's application should never under-perform because of \ac{vsnf} operations and (ii) the \ac{vsnf} placement must obey the \ac{tsp}'s security best-practices in terms of application security requirements, position in the network, operational mode (stateless or stateful \ac{vsnf}), and order with respect to the direction of the traffic. In the next section, we present the \ac{pess} mathematical model for the placement of \ac{vsnf} chains based on these criteria. }

\section{PESS \ac{vsnf} placement model}\label{sec:model}

\begin{figure}[!h]
	\begin{center}
		\includegraphics[width=0.48\textwidth]{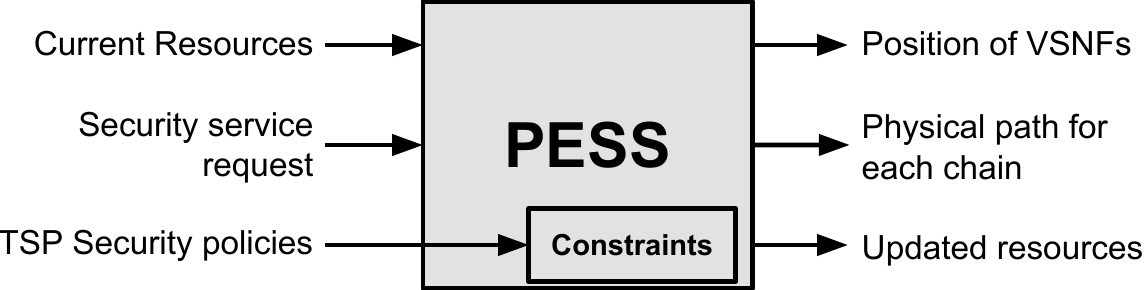}
		\caption{PESS placement model workflow.}
		\label{fig:PESS-chart}
	\end{center}
\end{figure}

The \ac{pess} model (Fig.~\ref{fig:PESS-chart}) is a mathematical model to progressively embed service requests, formed by one or multiple \ac{vsnf} chains, onto a physical network substrate by considering the available resources and realistic constraints.

PESS takes as input a model of the physical network including the current status of computing and network resources of servers and links, a security service request and the \ac{tsp}'s security policies (expressed in the form of constraints for PESS). The output of PESS is the mapping of the \acp{vsnf} onto the physical network (position of the \acp{vsnf} and one or more paths between them) and an updated model of the physical network taking into account the resources used to provision the service. The updated model is used as input for the next request.


\textbf{Physical network model.} We represent the physical network as a weighted graph $\mathcal{G} = (N,E)$, i.e. a graph where weights are assigned to nodes and edges.

\original{
Without loss of generality and to simplify the model, we assume that every node $i\in N$ is a NFVI-POP (Network Function Virtualization Infrastructure Point of Presence) characterized by the computing resources $\gamma_i\in\mathbb{N}^+$ expressed in CPU cycles/sec.
}

\revised{
Without loss of generality and to simplify the model, we assume that every node $i\in N$ is a NFVI-POP (Network Function Virtualization Infrastructure Point of Presence) \cite{nfvipop} consisting of a set of servers and a local network composed of routers and switches. Each node $i$ is characterized by the total computing resources of the servers $\gamma_i\in\mathbb{N}^+$ expressed in CPU cycles/sec.
}
A link $(k,l)\in E$ is a wired connection between two nodes $k$ and $l$ $\in N$. 
It is characterized by its capacity $\beta_{k,l}\in\mathbb{N}^+$ and its propagation delay $\lambda_{k,l}\in\mathbb{N}^+$. Both are expressed as positive integer numbers representing bandwidth (bits/sec) and latency (sec). 

\textit{Regions} in a physical network are defined as subsets of nodes sharing some high-level features. Examples of regions are: (i) a set of nodes in the TSP network providing the same cloud service (e.g. multimedia caching, data storage, etc.), or (ii) the set of egress nodes that connect the TSP network to the Internet (called \textit{border region} in the rest of this paper).

\textbf{Security service request.} We model a security service request as a set of independent weighted directed graphs:
$$\mathcal{G}_s = \{(U^c,U_{pairs}^c): c\in C_s\}$$
where $C_s$ is the set of unidirectional chains composing the service request.  
Each graph includes nodes and arcs. Nodes $U^c=A^c\cup V^c$ comprise user and remote applications ($A^c$, the endpoints of chain $c$) as well as a subset of all \acp{vsnf} ($V^c$). Each arc in $U_{pairs}^c$ delineates the order of traversing the \acp{vsnf} $\in V^c$ between endpoints in $A^c$.

Each chain $c\in C_s$ is characterized by its requirements in terms of minimum bandwidth $\beta^c$ and maximum latency $\lambda^c$. Each endpoint in $A^c$ is characterized by an identifier, which specifies where the endpoint must be placed in the physical network. The user application is characterized by the identifier of the physical node to which the user is attached (called \textit{ep1} in the rest of the paper). A remote application is characterized by the identifier of a region in the physical network (called \textit{EP2}). For instance, the border region if the endpoint represents a remote gaming server located outside the physical network. In this paper, $ep1$ and $EP2$ are referred to as \textit{physical endpoints} of the service request $\mathcal{G}_s$.\\
A \ac{vsnf} $u\in V^c$ is characterized by its requirements in terms of CPU units $\gamma_u$ expressed in CPU cycles/bit. $u$ is also characterized by the latency $\lambda^c_{i,u}$ it introduces in the dataplane to process a packet of chain $c$ on node $i$. As formalized in Eq. (\ref{eq:cpu_latency1}), this latency is a function of the residual computing capacity of the node $i$ where $u$ is placed, the computing requirements $\gamma_u$ of the \ac{vsnf}, the average packet size $\sigma^c$ of chain $c$ and the traffic load of the chain (whose upper bound is $\beta^c$). Finally, a \ac{vsnf} is characterized by its operational mode (either stateless or stateful) and by the identifier of a region in the physical network where it must be placed, if required by the \ac{tsp} security policies.

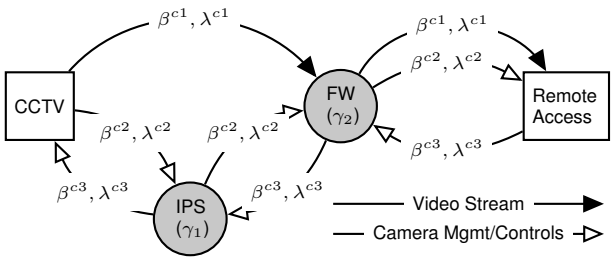
\begin{figure}[h!]
	\begin{center}
		\begin{tikzpicture}[every node/.style={font=\sffamily\scriptsize}, vnf/.style={circle,thick,draw,fill={rgb,255:red,200; green,200; blue,200}},app/.style={rectangle,thick,draw,font=\sffamily\scriptsize,minimum size=0.9cm}]
		
		\node[app] (1) {CCTV};
		\node[vnf,text width=0.38cm] (2) at (4,0) {FW\\($\gamma_2$)};
		\node[vnf,text width=0.38cm] (3) at (2,-1.5) {IPS\\($\gamma_1$)};
		\node[app,text width=0.90cm] (4) at (7,0) {Remote Access};
		
		\path[->,>=triangle 45,thick]
		(1) edge [bend left=50] node[midway,fill=white] {$\beta^{c1},\lambda^{c1}$} (2)
		(2) edge [bend left=60] node[fill=white] {$\beta^{c1},\lambda^{c1}$} (4);
		\path[every node/.style={font=\sffamily\scriptsize},->,>=open triangle 45,thick]
		(1) edge [bend left] node[fill=white] {$\beta^{c2},\lambda^{c2}$} (3)
		(3) edge [bend left] node[fill=white] {$\beta^{c2},\lambda^{c2}$} (2)
		(2) edge [bend left] node[fill=white] {$\beta^{c2},\lambda^{c2}$} (4);
		\path[every node/.style={font=\sffamily\scriptsize},->,>=open triangle 45,thick]
		(4) edge [bend left] node[fill=white] {$\beta^{c3},\lambda^{c3}$} (2)
		(2) edge [bend left] node[fill=white] {$\beta^{c3},\lambda^{c3}$} (3)
		(3) edge [bend left] node[fill=white] {$\beta^{c3},\lambda^{c3}$} (1);
		
		\draw[->,>=triangle 45,thick] (3.9,-1.3) -- (7.5,-1.3) node[midway,fill=white] {Video Stream};
		\draw[->,>=open triangle 45,thick] (3.9,-1.7) -- (7.5,-1.7	) node[midway,fill=white] {Camera Mgmt/Controls};
		\end{tikzpicture}
		\caption{Example of security service request for the CCTV system.}
		\label{fig:security_service_request}
	\end{center}
\end{figure}

\textbf{Illustrative example.} An example of a security service request for a CCTV system (see Table~\ref{tab:tspapps}) is represented in Fig.~\ref{fig:security_service_request}. The request in the example is composed of three chains (c1, c2, and c3), each one identified by the type of traffic and its direction.

Chain \textit{c1} is applied to the live video stream captured by the cameras and accessible over the Internet. The chain comprises a L3 firewall to ensure that the stream is only transmitted to authorized endpoints. As specified in Table~\ref{tab:tspapps}, the most relevant requirement in this case is the bandwidth ($\beta^{c1}$) which depends on the frame rate, frame size and video codec of the CCTV system. In this case, a deep inspection of the video stream packets (e.g., with an \ac{ips}) would not provide any additional protection but would possibly reduce the frame rate of the video streaming, thus compromising the detection of anomalous events. On the other hand, the bi-directional control/management traffic is inspected by the \ac{ips} and the firewall included in chains $c2$ and $c3$. Such \acp{vsnf} protect the CCTV system from attacks such as Mirai \cite{mirai} perpetrated through bots maliciously installed on Internet-connected devices, while the latency requirements $\lambda^{c2}$ and $\lambda^{c3}$ guarantee the responsiveness of the remote control of the CCTV cameras (pan, tilt, zoom, etc.).

\subsection{\ac{ilp} formulation}

\noindent\textbf{Definitions.} Let us first define two binary variables:
\begin{itemize}
	\item $x^c_{i,u}=1$ iff node $u\in U^c$ is mapped to $i\in N$.
	\item $y^c_{k,l,i,j,u,v}=1$ iff physical link $(k,l)\in E$ belongs to the path between nodes $i$ and $j$ to which $u,v\in U^c$ are mapped.
\end{itemize}

The residual capacity of a link, $\beta'_{k,l}$, is defined as the total amount of bandwidth available on link $(k,l)\in E$:
\begin{equation}\label{eq:residual_bandwidth}
\beta'_{k,l} = \beta_{k,l} - \sum_{\mathclap{\substack{c\in C,\ i,j\in N\\(u,v)\in U^c_{pairs}}}}\beta^c\cdot y^c_{k,l,i,j,u,v}
\end{equation}

thus, it is the nominal capacity of link $(k,l)$ minus the bandwidth required by the chains $c\in C$ already mapped on that link.

Similarly, the residual capacity of a node is defined as its nominal CPU capacity minus the computing resources used by the \acp{vsnf} $v$ instantiated on the node:
\begin{equation}\label{eq:residual_cpu}
\textstyle \gamma'_i = \gamma_i - \sum_{\{c\in C, u\in V^c\}} \gamma^c_u\cdot x^c_{i,u}
\end{equation}

\textbf{Problem formulation.} Given a physical network $\mathcal{G}$, for each security service request $\mathcal{G}_s$, find a suitable mapping of all its unidirectional chains on the physical network, which minimizes the physical resources of $\mathcal{G}$ expended to map $\mathcal{G}_s$, also known as the \textit{embedding cost}.\\
Hence, the solution of the problem is represented by a set of $x^c_{i,u}$ and $y^c_{k,l,i,j,u,v}$ such that the cumulative usage of physical resources for all the chains in $\mathcal{G}_s$ is minimized:

\begin{equation}\label{eq:objective_function}
\min\quad\sum_{\mathclap{\substack{c\in C_s,\ i,j\in N,\\(k,l)\in E,(u,v)\in U^c_{pairs}}}}b_{k,l}\cdot \beta^c\cdot y^c_{k,l,i,j,u,v}+\alpha\sum_{\mathclap{\substack{c\in C_s,i\in N, u\in V^c}}}c_{i}\cdot \gamma^c_u\cdot x^c_{i,u}
\end{equation}

Here, $\alpha$ is a factor that can be used to tune the relative weight of the cost components (we have used $\alpha=1$ for the experiments described in Section~\ref{sec:evaluation}).\\
$b_{k,l}$ and $c_{i}$ are the costs for allocating bandwidth and CPU: 
$$b_{k,l}=\frac{1}{\beta'_{k,l}+\delta}\quad c_{i}=\frac{1}{\gamma'_i+\delta}$$
They penalize nodes and links with less residual capacity with the aim to increase the chances of accommodating more security service requests on the given physical network. $\delta\longrightarrow 0$ is a small positive constant used to avoid dividing by zero in computing the value of the function.\\
\vspace{-4mm}
\subsection{Constraints}\label{sec:constraints}
\textbf{Routing Constraint} (\ref{eq:const_1}) ensures that each node $u\in U^c$ is mapped to exactly one physical node $i\in N$. With Constraint (\ref{eq:const_2}), a physical link $(k,l)$ can belong to a path between two nodes $i$ and $j$ for an arc $(u,v)\in U_{pairs}^c$ of chain $c\in C_s$ only if $u$ and $v$ are mapped to these nodes. Constraint  (\ref{eq:const_3}) ensures that the path created for arc $(u,v)$ starts at exactly one edge extending from node $i$ to where \ac{vsnf} (or start/endpoint) $u$ is mapped. Similarly, (\ref{eq:const_4}) ensures the correctness and the uniqueness of the final edges in the path. Constraints (\ref{eq:const_2}-\ref{eq:const_4}) can be easily linearized with standard techniques such as the ones presented in~\cite{Sherali2013}. Constraint (\ref{eq:const_5a}) is the classical \textit{flow conservation constraint}. That is, an outbound flow equals an inbound flow for each intermediate node $l$ (intermediate nodes cannot consume the flow). Together with Constraint (\ref{eq:const_5a}), Constraint (\ref{eq:const_5b}) prevents multiple incoming/outgoing links carrying traffic for a specific flow in the intermediate node $l$, i.e., we only consider unsplittable flows.
\begin{fleqn}[2pt]
	\begin{align}
	\begin{split}\label{eq:const_1}
	&\textstyle\sum_{\{i\in N\}} x^c_{i,u}=1\quad\forall c\in C_s,\forall u\in U^c
	\end{split}\\[2ex]
	\begin{split}\label{eq:const_2}
	& y^c_{k,l,i,j,u,v}\le x^c_{i,u}\cdot x^c_{j,v}\\
	&\forall c\in C_s, \forall i,j\in N, \forall (u,v)\in U^c_{pairs}, \forall (k,l)\in E 
	\end{split}\\[2ex]
	\begin{split}\label{eq:const_3}
	&\textstyle \sum_{\{(i,k)\in E, j\in N\}}y^c_{i,k,i,j,u,v}\cdot x^c_{i,u}\cdot x^c_{j,v}=1\\
	&\forall c\in C_s, \forall (u,v)\in U^c_{pairs}
	\end{split}\\[2ex]
	\begin{split}\label{eq:const_4}
	&\textstyle \sum_{\{(k,j)\in E, i\in N\}}y^c_{k,j,i,j,u,v}\cdot x^c_{i,u}\cdot x^c_{j,v}=1\\
	&\forall c\in C_s, \forall (u,v)\in U^c_{pairs}
	\end{split}
	\end{align}
\end{fleqn}

\begin{fleqn}[2pt]
	\begin{align}
	\begin{split}\label{eq:const_5a}
	&\sum_{\mathclap{\substack{k\in N\\(k,l)\in E}}}y^c_{k,l,i,j,u,v}=\sum_{\mathclap{\substack{m\in N\\(l,m)\in E}}}y^c_{l,m,i,j,u,v}
	\end{split}
	\end{align}
	\vspace{-6mm}
	\begin{align*}
	&\forall c\in C_s, \forall i,j\in N, \forall l\in N, l\ne i, l\ne j, \forall (u,v)\in U^c_{pairs} 
	\end{align*}
	\begin{align}
	\begin{split}\label{eq:const_5b}
	&\sum_{\mathclap{\substack{k\in N\\(k,l)\in E}}}y^c_{k,l,i,j,u,v}\leq 1
	\end{split}
	\end{align}
	\vspace{-6mm}
	\begin{align*}
	&\forall c\in C_s, \forall i,j\in N, \forall l\in N, l\ne i, l\ne j, \forall (u,v)\in U^c_{pairs} 
	\end{align*}
\end{fleqn}


\begin{table}[!h]
	\centering
	\scriptsize
	\caption{Glossary of symbols.}
	\vspace{-2mm}
	\label{tab:notations}
	\begin{tabular}{c}
		\textit{Sets}
	\end{tabular}
	\\
	\begin{tabular}{|L{1.2cm}|L{6.8cm}|}
		\hline
		\textit{N} & Set of physical nodes \\
		\hline
		\textit{E} & Set of physical links \\
		\hline
		\textit{$C$} & Set of all unidirectional chains already embedded in the network \\
		\hline
		\textit{$C_s$} & Set of all unidirectional chains in the service request $\mathcal{G}_s$\\
		\hline
		\textit{$U^c$} & Set of virtual nodes in the chain $c$ \\
		\hline
		$U^c_{pairs}$ & Set of unidirectional arcs in the chain $c$ \\
		\hline
		\textit{$A^c$} & Set of endpoints of the chain $c$. $A^c\subset U^c$ \\
		\hline
		\textit{$V^c$} & Set of \acp{vsnf} in the chain $c$. $V^c\subset U^c$ \\
		\hline
		\textit{$R_u$} & Region of $N$ where \ac{vsnf} $u$ must be placed ($region$ constraint) \\
		\hline
		\textit{$M$} & Region of $N$ where no \acp{vsnf} can be placed ($veto$ constraint) \\
		\hline
		\textit{$ep1, EP2$} & Physical endpoints of a service request. $ep1\in N, EP2\subset N$ \\
		\hline
		
	\end{tabular}
	\\
	\vspace{1mm}
	\begin{tabular}{c}
		\textit{Parameters}
	\end{tabular}
	\\
	\begin{tabular}{|L{0.9cm}|L{7.1cm}|}
		\hline
		\textit{$\gamma_i$} & Nominal computing resources of node $i$ (CPU cycles/sec)\\
		\hline
		\textit{$\gamma'_i$} & Residual computing resources of node $i$ (CPU cycles/sec) \\
		\hline
		\textit{$\gamma_u$} & CPU cycles required by $u$ to process one bit of a network packet (CPU cycles/bit)\\
		\hline
		\textit{$\gamma^c_u$} & Computing resources required by node $u$ of chain $c$ (CPU cycles/sec). $\gamma^c_u = \gamma_u\cdot \beta^c$\\
		\hline
		\textit{$\beta_{k,l}$} & Nominal capacity of link $(k,l)$ (bits/sec) \\
		\hline
		\textit{$\beta'_{k,l}$} & Residual capacity of link $(k,l)$ (bits/sec)\\
		\hline
		\textit{$\beta^c$} & Minimum bandwidth required by chain $c$  (bits/sec)\\
		\hline
		\textit{$\lambda_{k,l}$} & Propagation delay: the time spent by a packet to traverse the link $(k,l)$ (secs)\\
		\hline
		\revised{\textit{$\lambda^c_{k,l,i,j}$}} & \revised{Queuing delay: time spent by a packet of chain $c$ to traverse the network devices (routers and switches) in the local networks of adjacent nodes $k$ and $l$ (secs). $\lambda^c_{k,l,i,j}>0$ iff $k=i$ or $l=j$.}\\
		\hline
		\textit{$\lambda^c_{i,u}$} & Processing delay: the time spent by a packet to traverse \ac{vsnf} $u$ of chain $c$ placed on node $i$ (secs)\\
		\hline
		\textit{$\lambda^c$} & Maximum latency tolerated by chain $c$ (secs)\\
		\hline
		\textit{$\pi^c$} & Estimated latency between the \ac{tsp} network and the remote endpoint of chain $c$ (secs). $\pi^c=0$ if the endpoint belongs to the \ac{tsp} network.\\
		\hline
		\textit{$\sigma^c$} & Average packet size of chain $c$ (bits).\\
		\hline
		\textit{$b_{k,l}$} & Cost for allocating a unit of bandwidth on link \textit{(k,l)}\\
		\hline
		\textit{$c_{i}$} & Cost for allocating a unit of CPU on node \textit{i}\\
		\hline
	\end{tabular}
	\\
	\vspace{1mm}
	\begin{tabular}{c}
		\textit{Decision variables}
	\end{tabular}
	\\
	\begin{tabular}{|L{1.3cm}|L{6.7cm}|}
		\hline
		$x^c_{i,u}$ & Binary variable such that $x^c_{i,u}=1$ iff node $u\in U^c$ is mapped to $i\in N$ \\
		\hline
		$y^c_{k,l,i,j,u,v}$ & Binary variable such that $y^c_{k,l,i,j,u,v}=1$ iff physical link $(k,l)\in E$ belongs to the path between nodes $i$ and $j$ to which $u,v\in U^c$ are mapped\\
		\hline
	\end{tabular}
\end{table}

\textbf{Resource Constraints} (\ref{eq:const_8}-\ref{eq:const_9}) ensure that the resources consumed by a security service do not exceed the available bandwidth and computing capacities.
\begin{fleqn}[12pt]
	\begin{align}
	\begin{split}\label{eq:const_8}
	&\sum_{\mathclap{\substack{c\in C_s,\ i,j\in N\\(u,v)\in U^c_{pairs}}}}y^c_{k,l,i,j,u,v}\cdot \beta^c\le \beta'_{k,l}\quad \forall (k,l)\in E
	\end{split}\\[0.5ex]
	\begin{split}\label{eq:const_9}
	&\textstyle \sum_{\{c\in C_s, u\in V^c\}} x^c_{i,u}\cdot \gamma^c_u\le \gamma'_i\quad\forall i\in N
	\end{split}
	\end{align}
\end{fleqn}

\textbf{\ac{qos} Constraint} (\ref{eq:const_latency}) verifies that the requirements in terms of maximum end-to-end latency are met. It takes into consideration the propagation delay of physical links, the processing delay of \acp{vsnf} and the queuing delay through network devices. 
Note that the minimum bandwidth requirement is verified against the bandwidth resource Constraint (\ref{eq:const_8}).
\revised{
\begin{fleqn}[12pt]
	\begin{align}
	\begin{split}\label{eq:const_latency}
	\pi^c+\sum_{\mathclap{i\in N, u\in V^c}} x^c_{i,u}\cdot\lambda^c_{i,u}+\sum_{\mathclap{\substack{i,j\in N,(k,l)\in E\\(u,v)\in U^c_{pairs}}}}y^c_{k,l,i,j,u,v}\cdot (\lambda_{k,l}+\lambda^c_{k,l,i,j})\le \lambda^c&\\[-3.5ex]
	\forall c\in C_s\quad\qquad&\\[-3ex]
	\end{split}
	\end{align}
\end{fleqn}
}

$\pi^c$ is an estimation of the propagation delay between the \ac{tsp} network and the remote endpoint of chain $c$, in case the endpoint is outside the \ac{tsp} network. We assume that this value is independent from the \ac{tsp}'s network egress node. Clearly $\pi^c$ is $0$ for those chains whose remote endpoint is part of the \ac{tsp} network (e.g., a cloud data center managed by the \ac{tsp}). 

The processing delay $\lambda^c_{i,u}$ is the time spent by a packet to traverse \ac{vsnf} $u$ on physical node $i$. It contributes to the overall end-to-end delay of chain $c$ only if \ac{vsnf} $u$ is placed on node $i$ (i.e., $x^c_{i,u} = 1$).
$\lambda^c_{i,u}$ includes the time taken by the \ac{vsnf} to process the packet and the overhead of the virtualization technology (VMware, KVM, QEMU virtual machines, Docker containers, etc.). For simplicity, we do not model the delays due to the CPU scheduler operations  implemented on the physical node~\cite{8704949}. 
Based on the observations in \cite{GAO2018108,oljira-qos-aware}, \cite{7842188}, $\lambda^c_{i,u}$ is modeled as a convex function of the traffic load of the chain, and its value is computed by considering the impact of other \acp{vsnf} co-located on the same physical node.
\begin{equation}\label{eq:cpu_latency1}
\lambda^c_{i,u} = \frac{\gamma_u\cdot \sigma^c}{(\gamma'_i - \gamma_u\cdot\beta^c)+\delta}  = \frac{\gamma_u\cdot \sigma^c}{(\gamma'_i - \gamma^c_u)+\delta}
\end{equation}

In Eq. (\ref{eq:cpu_latency1}), $\gamma_u\cdot\sigma^c$ is the average amount of CPU cycles used by \ac{vsnf} $u$ to process a packet of chain $c$ (virtualization overhead included). The latency overhead caused by co-located \acp{vsnf} depends on the amount of computing resources of the node they use or, equivalently, on the residual computing resources of the node $\gamma'_i$. $\gamma_u\cdot\beta^c=\gamma^c_u$ is the amount of CPU cycles/sec used by \ac{vsnf} $u$ on node $i$, which depends on the traffic load of the chain.  
$\delta$ is a small positive constant used to avoid dividing by zero in the case that $u$ consumes all the residual computing resources of node $i$.

\original{
In our model, the propagation delay $\lambda_{k,l}$ is a fixed value (i.e., traffic-load independent) proportional to the length of the physical link $(k,l)$. 
}

\revised{	
The sum $\lambda_{k,l}+\lambda^c_{k,l,i,j}$ in Eq. (\ref{eq:const_latency}) is the total time spent by a packet travelling between two adjacent nodes $k$ and $l$. It includes the propagation delay $\lambda_{k,l}$, proportional to the distance between $k$ and $l$, and the queuing delay $\lambda^c_{k,l,i,j}$, proportional to the number of network devices (switches and routers) the packet traverses within the local networks of $k$ and $l$. The queuing delay is influenced by the buffer size of network devices' ports and by the traffic load \cite{silo}. For the sake of simplicity, we assume that the buffers are correctly dimensioned, i.e. no dropped packets due to buffer overflow. In addition, we estimate the queuing delay $\lambda^c_{k,l,i,j}$ as a traffic-load independent value; a function of the maximum queue capacity of the ports and of the \acp{vsnf} placement (hence a function of indices $k,l,i$ and $j$). Specifically, $\lambda^c_{k,l,i,j}>0$ if at least one \ac{vsnf} is mapped either on $k$ ($k=i$), or on $l$ ($l=j$), meaning that a packet of chain $c$ must traverse the local network of either $k$, or $l$ (or both) to reach the \acp{vsnf} running on the nodes' servers. Otherwise, the local networks of $k$ and $l$ are by-passed by the traffic of $c$, resulting in $\lambda^c_{k,l,i,j}=0$.
}

Constraint (\ref{eq:check_latency}) ensures that the current security service $\mathcal{G}_s$ does not compromise the end-to-end latency of chains $\hat{c}\in C$ in operational security services (also called \textit{operational chains} in the rest of the paper). 
\revised{
\begin{fleqn}[12pt]
	\begin{align}
	\begin{split}\label{eq:check_latency}
	\pi^{\hat{c}}+\sum_{\mathclap{i\in N, \hat{u}\in V^{\hat{c}}}} \bar{x}^{\hat{c}}_{i,\hat{u}}\cdot \lambda^{\hat{c}}_{i,\hat{u}}+\sum_{\mathclap{\substack{i,j\in N,(k,l)\in E\\(\hat{u},\hat{v})\in U^{\hat{c}}_{pairs}}}}\bar{y}^{\hat{c}}_{k,l,i,j,\hat{u},\hat{v}}\cdot (\lambda_{k,l}+\lambda^{\hat{c}}_{k,l,i,j})\le \lambda^{\hat{c}}&\\[-3.5ex]
	\forall {\hat{c}}\in C\quad\qquad&\\[-3ex]
	\end{split}
	\end{align}
\end{fleqn}
}
In Eq. (\ref{eq:check_latency}), $\bar{x}^{\hat{c}}$ and $\bar{y}^{\hat{c}}$ are the values of decision variables $x$ and $y$ computed for the placement of chain $\hat{c}$. $\lambda^{\hat{c}}_{i,\hat{u}}$ is the updated value of the processing delay introduced to the traffic of chain $\hat{c}$ by \ac{vsnf} $\hat{u}$ when running on node $i$. 

\begin{equation}\label{eq:cpu_latency_update}
\lambda^{\hat{c}}_{i,\hat{u}} = \frac{\gamma_{\hat{u}}\cdot \sigma^{\hat{c}}}{(\gamma'_i - \sum_{\{c\in C_s,u\in V^c\}}{x^c_{i,u}\cdot\gamma^c_u)}+\delta}
\end{equation}

In Eq. (\ref{eq:cpu_latency_update}), the value of $\lambda^{\hat{c}}_{i,\hat{u}}$ is updated by considering the computing resources consumed on node $i$ by \acp{vsnf} of the security service request $\mathcal{G}_s$. Approximation of Eq. (\ref{eq:cpu_latency_update}) can be achieved by using piecewise linearization techniques and Special-Ordered Set (SOS) variables and constraints available in most commercial solvers (e.g., \cite{gurobi_sos}).

\textbf{Security constraints} ensure that the \ac{tsp}'s security policies are applied. Specifically, Constraint (\ref{eq:const_stateful}) forces a subset $C'_s$ of the chains in the request to share the same \ac{vsnf} instance in case of stateful flow processing. 

\begin{equation}\label{eq:const_stateful}
\textstyle  x^{c_1}_{u,i}=x^{c_2}_{u,i}\quad \forall c_1,c_2\in C'_s\subset C_s, i\in N, u\in V^c 
\end{equation}

Constraint (\ref{eq:const_region}) forces the algorithm to place the \ac{vsnf} $u\in V^c$ in a specific region of the network defined as a subset of nodes $R_u\subset N$.

\begin{equation}\label{eq:const_region}
\textstyle \sum_{\{i\in R_u\}} x^c_{i,u}=1\quad  R_u\subset N, R_u\neq\emptyset, u\in V^c
\end{equation}
We use Constraint (\ref{eq:const_region}) to enforce the security close to the user by placing \acp{vsnf} on \textit{ep1} ($R_u=\{ep1\}$), or to protect a portion of the \ac{tsp}'s network, such as the border region or a distributed data center ($R_u=EP2$) from potentially malicious user traffic. {\color{diffcolor} Furthermore, Constraint (\ref{eq:const_region}) can be used to place a \ac{vsnf} on a physical node with special hardware characteristics (e.g., hardware acceleration for encryption).} Similarly, the \textit{veto} Constraint (\ref{eq:const_veto}) can be used to prevent the placement of any \acp{vsnf} on a pre-defined subset of nodes $M\subset N$. A \ac{tsp} may choose to do this to protect specific nodes (called \textit{veto nodes}) that host sensitive data or critical functions from user traffic.
\begin{equation}\label{eq:const_veto}
\textstyle \sum_{\{i\in M, u\in V^c\}} x^c_{i,u}=0\quad  \forall c\in C_s, M\subset N, M\neq\emptyset
\end{equation}	
Finally, for each chain $c\in C_s$, the correct order of \acp{vsnf} in $V^c$ is ensured by Constraints (\ref{eq:const_1}-\ref{eq:const_5b}), plus Constraint (\ref{eq:const_region}) applied to user and remote applications $u\in A^c$ with $R_u=\{ep1\}$ and $R_u=EP2$ respectively.
Note that, the order can be specified per application (chain), as different applications may require the same \acp{vsnf} but in different order. 
	
These four security constraints enable fulfillment of the security policies/practices defined by the \ac{tsp} e.g. the order in which the \acp{vsnf} are executed, the position of the \acp{vsnf} in the network, and the operational mode of \acp{vsnf} (either stateful or stateless).
\section{The PESS Heuristic algorithm}\label{sec:implementation}

The embedding problem presented in Section~\ref{sec:model} has been solved using a commercial solver. However, given the complexity of the \ac{ilp} model, the solver is unable to produce solutions in an acceptable time frame, as required for dynamic scenarios such as the ones under study. For this reason, we have also implemented a heuristic algorithm to find near optimal solutions in much shorter time. 



The logic behind the PESS heuristic is based on assuring that Constraints (\ref{eq:const_1}-\ref{eq:const_veto}) are applied in an efficient manner. In particular, the security constraint (\ref{eq:const_stateful}) ensures that a stateful \ac{vsnf} specified in two or more chains in the same service request $\mathcal{G}_s$ is placed on the same node. However, as different chains might share more than one stateful \ac{vsnf} (possibly in a different order), the correct placement of a multi-chain security service request may become a computationally expensive operation.
For this reason, given a path between $ep1$ and one of the nodes $ep2\in EP2$, the heuristic places all the \acp{vsnf} specified in $\mathcal{G}_s$ on a maximum of three nodes of the path with the following strategy: (i) place each region-specific \ac{vsnf} $u\in V^c$ ($R_u\neq\emptyset$) either on $ep1$ or on $ep2\in EP2$ depending on $R_u$ (i.e., either $R_u=\{ep1\}$ or $R_u=EP2$), (ii) place all the other \acp{vsnf} in $\mathcal{G}_s$ on the node with the highest residual capacity in the path to minimize the embedding cost (Eq. \ref{eq:objective_function}). 
 
 
The solution is obtained by selecting the candidate path between \textit{ep1} and \textit{EP2} where the embedding of all the chains in $\mathcal{G}_s$ fulfills the constraints described in Section~\ref{sec:model} at the lowest cost, as computed with the objective function (Eq. \ref{eq:objective_function}).



\begin{algorithm} 
	\caption{the PESS algorithm.}\label{alg:pess_algorithm}
	\begin{algorithmic}[1]
		\renewcommand{\algorithmicrequire}{\textbf{Input:}}
		\renewcommand{\algorithmicensure}{\textbf{Output:}}
		\Require Physical network substrate ($\mathcal{G}$), security service request ($\mathcal{G}_s$), set of active chains in the network ($C$)
		\Ensure The mapping of the security service onto the physical substrate (\textit{solution}). \textit{None} if no feasible mappings are found.
		\Procedure{PESS}{$\mathcal{G}$, $\mathcal{G}_s$, $C$}
		\State $\bar{\beta}\gets\sum_{\{c\in C_s\}}\beta^c$\Comment{total required bandwidth}
		\State $\bar{\gamma}\gets\sum_{\{c\in C_s, u\in V^c\}}\gamma^c_u$\Comment{total required CPU}
		\State $P=\{p_{[ep1,ep]}\mid ep\in EP2\}\gets \Call{Dijkstra}{ep1,EP2,\bar{\beta}}$
		\If {$P=\emptyset$}
		\State \textbf{return None}
		\EndIf
		\State $S=\{s_{[ep1,ep]}\mid ep\in EP2\}\gets \Call{Embed}{P,\bar{\beta},\bar{\gamma}}$
		\State $N_S\gets \{i\in s\mid s\in S\}$\Comment{physical nodes in the initial solutions}
		\State $E\gets \{i\in N\mid i\notin N_S\cup M,\ \gamma'_i>\gamma'_j\ \forall j\in N_S\}$
		\State $\bar{s}_{[ep1,ep2]}\gets \operatorname*{argmin}\limits_{s\in S} cost(s)$\Comment{best initial solution}
		\State $P_1\gets \Call{Dijkstra}{ep1,E,\bar{\beta}}$
		\State $P_2\gets \Call{Dijkstra}{ep2,E,\bar{\beta}}$
		\State $S\gets S\cup \Call{Embed}{P_1\cup P_2,\bar{\beta},\bar{\gamma}}$\Comment{expanded solution set}
		\State $solution\gets$\textbf{None}
		\State $S\gets \Call{SortedDecreasingCost}{S}$
		\ForAll {$cs\in S$}
		\If {$\Call{LatencyOpChains}{\mathcal{G},C,cs}$ \textbf{is True}}
		\State $solution\gets cs$
		\State \textbf{break}
		\EndIf
		\EndFor
		\If {$solution$ \textbf{is None}}
		\State \textbf{return None}
		\EndIf
		\State $\Call{UpdateResources}{solution,\mathcal{G}}$
		\State $\Call{StoreSolution}{\mathcal{G},C,solution}$
		\State \textbf{return} $solution$
		\EndProcedure
	\end{algorithmic}
\end{algorithm}

\textbf{Initial solution.} The embedding process starts at line 4 in Algorithm \ref{alg:pess_algorithm} with a greedy approach based on the Dijkstra's algorithm. At this stage, we compute the shortest path tree between the two endpoints $ep1$ and $EP2$ using the residual bandwidth as link weight computed as $b_{k,l}\cdot \beta^c$ in Eq. (\ref{eq:objective_function}). The Dijkstra algorithm stops when all the nodes $ep2\in EP2$ are marked as \textit{visited}, i.e. before building the whole tree of paths. For each path between \textit{ep1} and \textit{EP2}, the algorithm places the \acp{vsnf} in the chains according to the aforementioned strategy, the order of the \acp{vsnf} as specified in the service request, the latency Constraint (\ref{eq:const_latency}), and the security Constraints (\ref{eq:const_stateful}-\ref{eq:const_veto}) (line 8).
The output of this first step is a set of candidate solutions $S$ with different embedding costs. $S$ is passed as input to the next step. 

\textbf{Expanded solution set.} The algorithm now evaluates whether high-capacity nodes not included in the initial solution set $S$ can be used to build new solutions with lower embedding cost. Hence, given the initial set of solutions $S$, the algorithm identifies the physical nodes in the network with these two properties (set $E$ defined at line 10): (i) not included in the initial set of solutions $S$ nor \textit{veto} nodes, and (ii) higher computing capacity with respect to the nodes included in $S$. The algorithm then computes the shortest path tree twice, once from $ep1$ towards $E$ and once from $ep2\in EP2$ towards $E$ (lines 12-13), where $\{ep1,ep2\}$ are the physical endpoints of the solution in $S$ with the lowest embedding cost (line 11). 

The resulting subpaths are joined to form a new set of paths between \textit{ep1} and \textit{ep2}. Afterwards, the algorithm performs the placement of the \acp{vsnf} on each of the new paths with the strategy described earlier in this section. The feasible solutions are added to the initial set $S$ (line 14).

The set of candidate solutions is sorted in descending value of embedding cost (line 16). The first one that satisfies Constraint  (\ref{eq:check_latency}) is the accepted solution (lines 18-19).
Finally, the algorithm updates the values of $\gamma'_i$ and $\beta'_i$ by removing the resources consumed with the accepted solution and stores the mapping of its chains in the set $C$ that records all the active chains in the network (lines 26-27).

\textbf{Latency of operational chains.}
Given a candidate solution $cs\in S$, function $\Call{LatencyOpChains}{}$ is invoked to verify whether embedding $cs$ compromises the end-to-end latency of operational chains (line 18 in Algorithm \ref{alg:pess_algorithm}). Instead of verifying the inequality in Eq. (\ref{eq:check_latency}) for each operational chain, $\Call{LatencyOpChains}{}$ implements a heuristic approach, which reduces the time complexity of this operation from $O(n)$, with $n$ the number of operational chains, to $O(1)$.

Each time a chain $c\in C$ becomes operational, the algorithm computes $\langle\gamma\rangle^c$, a threshold value obtained from Eq. (\ref{eq:const_latency}) and (\ref{eq:cpu_latency1}) as follows:
\revised{
	\begin{fleqn}[12pt]
		\begin{align}
		\begin{split}\label{eq:average_residual_cpu}
		\langle\gamma\rangle^c = \frac{\sum\limits_{i\in N, u\in V^c} \bar{x}^c_{i,u}\cdot\gamma_u\cdot\sigma^c}{\lambda^c-\pi^c-\sum\limits_{\makebox[0pt]{$\scriptstyle \substack{i,j\in N,(k,l)\in E\\(u,v)\in U^c_{pairs}}$}}\bar{y}^c_{k,l,i,j,u,v}\cdot (\lambda_{k,l}+\lambda^c_{k,l,i,j})}-\delta&\\[-2ex]
		\end{split}
		\end{align}
	\end{fleqn}
}

In Eq. (\ref{eq:average_residual_cpu}), $\bar{x}$ and $\bar{y}$ are the values of decision variables $x$ and $y$ used to embed $c$.
$\langle\gamma\rangle^c$ estimates the minimum average residual computing capacity necessary to satisfy the inequality in Eq. (\ref{eq:const_latency}). Therefore, the algorithm records and monitors those operational chains with the highest values of $\langle\gamma\rangle^c$ to establish whether a candidate solution is feasible or not, as inequality in Eq. (\ref{eq:const_latency}) is violated earlier for such chains than for the others.

The algorithm stores one operational chain per physical node in a data structure, i.e. the chain with the highest value of $\langle\gamma\rangle^c$ with at least one \ac{vsnf} mapped on that node. Hence, given the physical nodes mapped in the candidate solution $cs$, the algorithm computes Eq. (\ref{eq:check_latency}) only for the operational chains in the data structure linked to such nodes by using the values of variables $x$ and $y$ of solution $cs$. If the inequality is not satisfied for one of those chains, $cs$ is rejected.

As the maximum number of physical nodes used to provision a security service is three (\textit{ep1} and \textit{EP2} to fulfill the region constraint and the node with the highest residual capacity in the path), the worst-case time complexity of this process is $O(1)$, thus constant in the number of operational chains and with respect to the size of the network.
Therefore, the overall time complexity of the \ac{pess} heuristic is $O(|E|+|N|\log(|N|))$, i.e., the worst-case time complexity of the Dijkstra's algorithm.

\section{Evaluation}\label{sec:evaluation}
We first assess the PESS heuristic by comparing its solutions against the optimal embeddings as computed by a commercial solver (Gurobi~\cite{gurobi}). 
We then prove the benefits of the proposed application-aware approach against the baseline (the application-agnostic approach adopted, for instance, in \cite{7592416}), in which security services are provided without taking into account the specific requirements of applications. We finally analyze the scalability of PESS by measuring the average embedding time on different network sizes.

\subsection{Test configuration}\label{sec:test_config}
The \ac{pess} heuristic has been implemented as a single-threaded Python program, while the ILP model formalized in Section \ref{sec:model} has been implemented with the Gurobi Python API version 7.5~\cite{gurobi_api}. All experiments are performed on a server-class computer equipped with 2 Intel Xeon Silver 4110 CPUs (16 cores each running at 2.1\,GHz) and 64\,GB of RAM. 
\subsubsection{Topology} 
The simulations are performed on synthetic topologies randomly generated based on the Barab\'asi-Albert model~\cite{barabasi}. We generate topologies of different sizes and densities to evaluate the performance of the \ac{pess} heuristic in a variety of generic network scenarios.

\original{
We also use the model of a real network; the Italian education and research network (consortium GARR~\cite{garr-whitepaper}) represented in \cite{garr-topology}. In addition to the actual view of the physical topology, \cite{garr-topology} also provides the specification of the egress nodes, i.e. the nodes that connect the GARR network to the Internet and that compose the \textit{border} region in our evaluation (nodes \textit{FI1, MI2, PD2, RM2} and \textit{TO1}, as indicated in \cite{garr-topology}). \cite{garr-backbone} specifies the nominal capacity of all the links in the GARR network.

As we have no information related to data center distribution in the GARR network, we have assumed one NFVI-POP for each node of the network with computing capacity of 32x2.1\,GHz (a 32-core CPU running at 2.1\,GHz).\\
In the rest of the evaluation, we refer to this topology as the \textit{reference network}.
}

\revised{We also validate \ac{pess} with two realistic network models. One is the Stanford University backbone \cite{headerspace}, a medium-scale campus network consisting of 46 links, 14 operational zone routers, 10 Ethernet switches, and 2 border routers connecting the University to the Internet. We assume one NFVI-POP for each network device and 10\,Gbps links. The second model is the Italian education and research network (consortium GARR~\cite{garr-whitepaper}). The GARR network covers the entire Italian national territory (\cite{garr-topology}), comprising 83 links and 46 nodes. Along with the actual view of the physical topology, \cite{garr-topology} provides the specification of the egress nodes, i.e. the nodes that connect the GARR network to the Internet and that compose the \textit{border} region in our evaluation (nodes \textit{FI1, MI2, PD2, RM2} and \textit{TO1}, as indicated in \cite{garr-topology}). As we have no information related to data center distribution in the GARR network, we have assumed one NFVI-POP for each node. In addition, we set the nominal capacity of the links $\beta_{k,l}$ using the values specified in \cite{garr-backbone}.
	
Given the relatively small size of the Stanford network, we assume no propagation delay between its nodes, i.e. $\lambda_{k,l}=0\ \forall k,l$. For the other networks, random and GARR, we compute the propagation delay of each link with the following formula:
$$\lambda_{k,l} = \frac{d_{k,l}\cdot r\_index}{C}$$
where $r\_index=1.5$ is an approximation of the refractive index of optical fibers, $C\simeq 3\cdot 10^8\ m/s$ is the speed of light in the vacuum and $d_{k,l}$ is the distance between two nodes $k$ and $l$. In the case of random networks, $d_{k,l}$ is a random positive value ranging from 10 to 100\,Km, while for the GARR network $d_{k,l}$ is computed by approximating the coordinates of the nodes based on the information available on the web site.

As introduced in Section \ref{sec:constraints}, we estimate the worst-case queuing delay $\lambda^c_{k,l,i,j}$ as a traffic-load independent value using the queue capacity of switch ports reported in \cite{silo} ($80\,\mu s$ for 10\,Gbps ports with a 100\,KB buffer). Specifically, we assume a three-tier local network at each node of GARR and random topologies, resulting in a maximum of $12\times80\,\mu s$ queuing delay introduced at each node. This reflects the maximum queuing delay experienced by each packet crossing a node to be processed by one or more \acp{vsnf} mapped on the node, which involves traversing three network devices (hence, six 10\,Gbps ports) to reach the servers where the \acp{vsnf} are running, and traversing three network devices before leaving the node (six further 10\,Gbps ports).   

For the campus scenario, implemented using the Stanford University topology, we instead assume only one network device per node; the device specified in the network topology. Hence, the maximum queuing latency for a packet crossing a Stanford node is  $4\times80\,\mu s$. 

For each node of the three evaluation scenarios we assume one server with computing capacity of 32x2.1\,GHz (a 32-core CPU running at 2.1\,GHz). 
}

\subsubsection{Security service requests}\label{sec:service_requests}
As introduced in Section \ref{sec:model}, a security service request is configured by the \ac{tsp} to provision security for user applications (see the CCTV example in Section  \ref{sec:model}). For evaluation purposes, we automatically generate requests composed of a random number of chains, ranging between 1 and 5. Each chain comprises a random subset of \acp{vsnf} from the list presented in Table \ref{tab:vsnf_cpu_requirements}, with a maximum of 3 \acp{vsnf} per chain
{\color{diffcolor}
(i.e., up to 15 \acp{vsnf} per user application). Based on the use case scenarios illustrated in Sections \ref{sec:motivation} and \ref{sec:model} (web browsing, online gaming, CCTV system), we believe these are reasonable values.
}

\begin{table}
	\centering
	\scriptsize
	\caption{CPU requirements for some \ac{vsnf} implementations.}
	\label{tab:vsnf_cpu_requirements}
	\begin{threeparttable}
		\begin{tabular}{V{2}>{\bfseries}l|c|c V{2}}
			\hlineB{2}
			\rowcolor{mygray}
			\textbf{\ac{vsnf}} & \textbf{Virtualization} & \begin{tabular}{@{}c@{}}{\boldmath $\gamma_u$}\\ \textbf{(cycles/bit)}\tnote{1}\end{tabular}\\
			\hlineB{2}
			Snort IDS/IPS & VirtualBox &9.5~\cite{SHAH2018157} \\
			\hline
			Suricata IDS/IPS & VirtualBox &8.2~\cite{SHAH2018157} \\
			\hline
			\begin{tabular}{@{}c@{}}OpenVPN with AES-NI\end{tabular}& KVM/QEMU &31~\cite{7973470}\\
			\hline
			\begin{tabular}{@{}c@{}}strongSwan with AES-NI\end{tabular}& KVM/QEMU &16~\cite{7973470}\\
			\hline
			\begin{tabular}{@{}c@{}}Fortigate-VM NGFW\end{tabular} & \begin{tabular}{@{}c@{}}FortiOS \end{tabular}& 9~\cite{fortigate}\\
			\hline
			\begin{tabular}{@{}c@{}}Fortigate-VM SSL VPN\end{tabular} & \begin{tabular}{@{}c@{}}FortiOS \end{tabular}& 13.6~\cite{fortigate}\\
			\hline
			\begin{tabular}{@{}c@{}}Fortigate-VM IPSec VPN\end{tabular} & \begin{tabular}{@{}c@{}}FortiOS \end{tabular}& 14.5~\cite{fortigate}\\
			\hline
			\begin{tabular}{@{}c@{}}Fortigate-VM Threat protection\end{tabular} & \begin{tabular}{@{}c@{}}FortiOS \end{tabular}& 11.3~\cite{fortigate}\\
			\hline
			\begin{tabular}{@{}c@{}}Cisco ASAv Stateful IDS\end{tabular} & \begin{tabular}{@{}c@{}}VMware ESX/ESXi \end{tabular}& 4.2~\cite{cisco_asav}\\
			\hline
			\begin{tabular}{@{}c@{}}Cisco ASAv AES VPN\end{tabular} & \begin{tabular}{@{}c@{}}VMware ESX/ESXi \end{tabular}& 6.9~\cite{cisco_asav}\\
			\hline
			Juniper vSRX FW & \begin{tabular}{@{}c@{}}VMware VMXNET3 \end{tabular}& 2.3~\cite{juniper_vsrx}\\
			\hline
			Juniper vSRX IPS & \begin{tabular}{@{}c@{}}VMware VMXNET3 \end{tabular}& 2.4~\cite{juniper_vsrx}\\
			\hline
			\begin{tabular}{@{}c@{}}Juniper vSRX AppMonitor\end{tabular}& \begin{tabular}{@{}c@{}}VMware VMXNET3 \end{tabular}& 1.5~\cite{juniper_vsrx}\\ \hlineB{2}
			
		\end{tabular}
		\begin{tablenotes}
			\item[1] $\gamma_u$=(CPU clock)*(CPU usage)/Throughput. CPU usage is set to 1 (i.e. 100\%) when the value is not specified.
		\end{tablenotes}
	\end{threeparttable}
\end{table}

The CPU requirements for the VSNFs are presented in Table \ref{tab:vsnf_cpu_requirements}. It should be noted that the values of $\gamma_u$ (cycle/bit) reported in Table \ref{tab:vsnf_cpu_requirements} are estimated based on the results of experiments reported in scientific papers or product datasheets and obtained under optimal conditions, with only one \ac{vsnf} running at a time. The impact on the network traffic caused by concurrent \acp{vsnf} running on the same node are estimated with Eq. (\ref{eq:cpu_latency1}) and (\ref{eq:cpu_latency_update}). 
These values of $\gamma_u$ have been used to perform the evaluation tests described in the remainder of this section, with the aim of enabling interested readers to replicate the experiments in similar conditions. However, we also obtained comparable results using random values.

\newcommand{\PreserveBackslash}[1]{\let\temp=\\#1\let\\=\temp}
\newcolumntype{C}[1]{>{\PreserveBackslash\centering}p{#1}}
\revised{
	\begin{table}
		\centering
		\scriptsize
		\renewcommand{\arraystretch}{1.1}
		\caption{\revised{Comparison between the PESS heuristic and the PESS ILP-based algorithm on three network scenarios.}}
		\label{tab:heuris_solver}
		\begin{threeparttable}
			\begin{tabular}{V{2}>{\bfseries}l V{2} c V{2} C{1cm}|C{1cm} V{2}}
				\hlineB{2}
				\rowcolor{mygray}
				\textbf{Network model} & \begin{tabular}{@{}c@{}}\textbf{Heuristic embedding}\\ \textbf{cost overhead}\tnote{1}\end{tabular} & \multicolumn{2}{c V{2}}{\begin{tabular}{@{}c@{}}\textbf{Average Time (sec)}\\ \textbf{Heuristic}\ \qquad \textbf{Solver}\ \ \ \end{tabular}} \\
				\hlineB{2}
				\textbf{Random}& 0.06\% & 0.002 & 150 \\ \hline
				\textbf{Stanford}& 0.07\% & 0.003 & 700 \\ \hline
				\textbf{GARR}& 0.5\% & 0.003 & 1500 \\ \hlineB{2}
			\end{tabular}
			\begin{tablenotes}
				\item[1]Average overhead with respect to the solver embedding cost.
			\end{tablenotes}
		\end{threeparttable}
	\end{table}
}

\subsection{Comparison between solver and heuristic}\label{sec:performance_comparison}
\textbf{Methodology.} In this experiment, we compare the PESS ILP-based algorithm implemented with the solver and the PESS heuristic \revised{on the Stanford and GARR network models}, and on Barab{\'a}si-Albert random topologies with 20 nodes and 36 links.

The security service requests are generated using a Poisson process with exponential distribution of inter-arrival and holding times. Once a service expires, the resources allocated to it are released.

We start by simulating the processing of $10^5$ service requests using the PESS heuristic. Once a stable network utilization (load) is reached, we save the subsequent service requests along with the network state and the heuristic solution. In a second stage, we run the solver to compute the optimal solution for each of the requests saved in the previous stage and we compare the results with the recorded heuristic solutions. This process is repeated with values of network load ranging between 1000 and 20000 Erlang.

\textbf{Metrics.} (i) Heuristic embedding cost overhead over optimal solutions and (ii) embedding time.

\textbf{Discussion.}  As explained in Section~\ref{sec:implementation}, the PESS heuristic places all the chains of a service request on a single path to efficiently guarantee that the \ac{qos} Constraint (\ref{eq:const_latency}) and the region Constraint (\ref{eq:const_region}) are respected. Once the path is found, the heuristic places the \acp{vsnf} of all the chains on a maximum of three nodes in the chosen path: the one with the highest residual computing capacity and the ones specified with the region constraint (if any). Such implementation choices reduce the solution space in case of requests with multiple chains and \acp{vsnf}. On the other hand, Constraints (\ref{eq:const_latency}) and (\ref{eq:const_region}) also narrow down the solution space for the solver, often resulting in single-path optimal solutions. As a result, \revised{we measure a marginal embedding cost overhead of the heuristic solutions with respect to the optimal solutions on all three evaluation scenarios (see Table \ref{tab:heuris_solver}).}

\revised{It is worth analyzing the reason behind nearly one order of magnitude difference between the GARR topology and the other two network scenarios.} When the initial solution is computed, the heuristic algorithm selects the endpoint $ep2\in EP2$ to further explore the solution space, thus excluding the other endpoints in $EP2$ (line 11 in Algorithm \ref{alg:pess_algorithm}). This strategy improves the scalability of the heuristic in case of large endpoint sets $EP2$, at the cost of slightly reducing the quality of the solutions.

In this regard, on the GARR network the border region is used as endpoint $EP2$ for 80\% of the requests, to simulate a real-world \ac{tsp} network where most of the traffic is directed towards the Internet. Hence, good solutions involving four of the five nodes in the border are not considered during the second stage of the heuristic, possibly leading to less accurate solutions. \revised{Conversely, a border region of only two nodes is defined in the Stanford topology (the two border routers), while no special regions at all are configured for the random networks (thus, always $|EP2|=1$), resulting in more precise embeddings.

As reported in Table \ref{tab:heuris_solver}, the embedding time measured for the heuristic is 3\,ms, on average, with the Stanford and GARR topologies, and below 3\,ms, on average, with the random topologies. In contrast, the solver takes between 150 and 1500\,s, on average, to find the optimal solutions on the three network scenarios. Please note that, the results related to the GARR network are limited to service requests with less than 10 \acp{vsnf}. Due to the size of the GARR topology (46 nodes and 83 links), above this threshold the solver runs out of memory and it is terminated by the operating system.}

\revised{
\begin{figure*}[!h]
	\centering
	\includegraphics[width=1\textwidth]{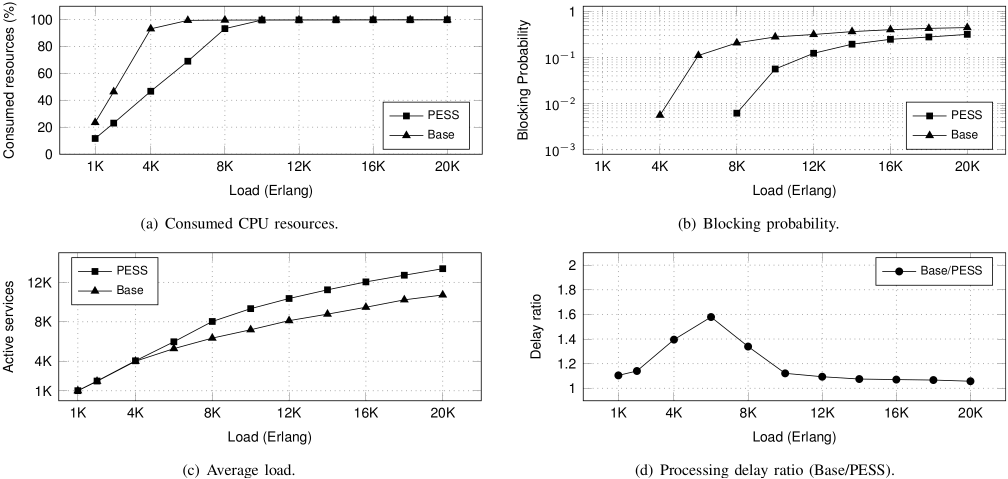}
	\caption{Comparison between the baseline (Base) and the PESS approaches on random networks (20 nodes and 36 links).}
	\label{fig:app_base_random}
	\vspace{-1em}
\end{figure*}
\begin{figure*}[!h]
	\centering
	\includegraphics[width=1\textwidth]{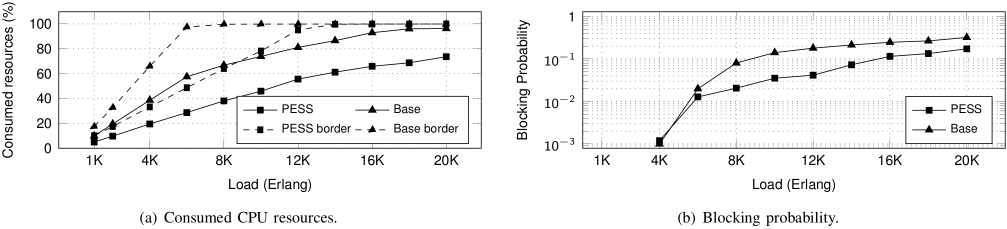}
	\caption{Comparison between the baseline (Base) and the PESS approaches on the GARR network.}
	\label{fig:app_base_garr}
	\vspace{-1em}
\end{figure*}
\begin{figure*}[!h]
	\centering
	\includegraphics[width=1\textwidth]{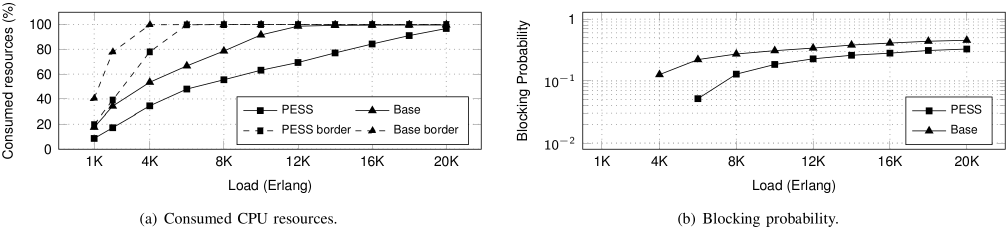}
	\caption{Comparison between the baseline (Base) and the PESS approaches on the Stanford backbone network.}
	\label{fig:app_base_stanford}
	\vspace{-1em}
\end{figure*}
}

\begin{figure*}[!h]
	\centering
	\includegraphics[width=1\textwidth]{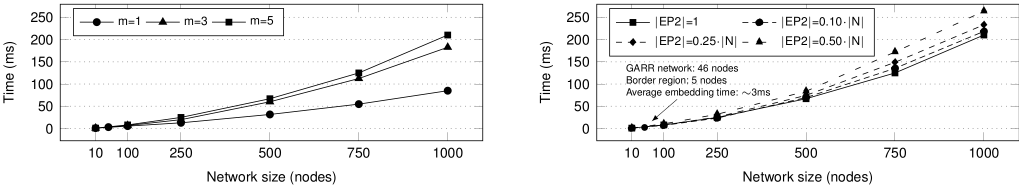}
	\caption{Heuristic execution time as a function of the number of physical nodes. The plot on the left reports the results with $m=1,3,5$ in the Barab{\'a}si-Albert model when generating random graphs, while the one on the right side shows the measurements with fixed $m=5$ at different sizes of endpoint $EP2$.}
	\label{fig:scalability-plots}
	\vspace{-1em}
\end{figure*}

\subsection{\ac{pess} vs application-agnostic provisioning}\label{sec:app-base-comparison}
\textbf{Methodology.}  We start two experiments in parallel using two identical copies of the same physical network graph. At each iteration, we generate a service request with application-specific \ac{qos} and security requirements. In \textit{Experiment 1}, the security service is provisioned on one copy of the network with the \ac{pess} heuristic. In \textit{Experiment 2}, the service is provisioned on the second copy of the network by simulating the standard approach (adopted, for instance, in \cite{7592416} and used in this test as baseline), where two application-agnostic chains of \acp{vsnf} (one for each direction of the traffic) are applied to the user traffic to fulfill all the security requirements regardless of the specific needs of the applications. At the end of each iteration, the two copies of the network are updated according to the resources consumed by the respective provisioning approach.

As in the previous experiment, security service requests are generated using a Poisson process with exponential distribution of inter-arrival and holding times. We run $10^5$ iterations, starting to collect statistics after the first $8\cdot 10^4$ requests (once a stable network load is reached). The two parallel experiments are repeated with different network load values.

\textbf{Metrics.} Blocking probability, consumption of computing resources, end-to-end latency of the chains and number of active services in the network.

\textbf{Discussion.} Fig. \ref{fig:app_base_random} compares the performance of the \ac{pess} application-aware service provisioning algorithm (\textit{PESS} in the figure) and the baseline approach (\textit{Base}) on random networks. The experimental results are plotted as functions of the network load, which is expressed in terms of the average number of security service requests in the network (Erlang).

The efficient usage of the computing resources reported in Fig. \ref{fig:app_base_random}(a) is a major benefit of the application-aware provisioning  mechanism proposed in this work. In particular, \ac{pess} avoids inefficiencies, such as a high bandwidth video stream being processed by a high resource demanding \ac{ips} (see the CCTV example in Section \ref{sec:model}), ultimately leading to a lower blocking probability and to a higher number of active services in the network, as shown in Fig. \ref{fig:app_base_random}(b) and \ref{fig:app_base_random}(c) respectively.

The benefits of \ac{pess} in terms of reduced end-to-end latency are reported in Fig. \ref{fig:app_base_random}(d). The plot illustrates the ratio between the average end-to-end latency of the chains in \textit{Experiment 2} (Baseline), and the average end-to-end latency of the chains in \textit{Experiment 1} (\ac{pess}). \revised{At low loads, when the nodes in the networks of both experiments are only partially busy, the value of this ratio is between 1.1 and 1.4. In other words, under typical operational conditions, the average end-to-end latency of chains provisioned with our approach is 10-40\% lower than the baseline.} Moreover, when the nodes in \textit{Experiment 2} are heavily loaded, the processing delay introduced by busy nodes becomes very high, as modeled with Eq. (\ref{eq:const_latency}). This phenomenon produces high ratios, represented by the spike in the plot, which gradually decrease at high loads when the nodes in the network of \textit{Experiment 1} also become fully loaded.

Fig. \ref{fig:app_base_garr} reports the results of the simulations performed with the GARR network. In this case, we are particularly interested in observing the behavior of our approach in the presence of a critical region (from the security viewpoint) such as the border of the network. In order to analyse this, we empirically configure the random generator of service requests to generate 80\% of requests directed towards the Internet (i.e., crossing the border of the network). In Fig. \ref{fig:app_base_garr}(b), it can be noted that both PESS and the baseline have similar blocking probability at low loads (below 6000). This is a consequence of the bandwidth usage on links towards the border region, which is almost always identical for \textit{Experiment 1} and \textit{Experiment 2}. The two curves start diverging at load 6000, i.e. when the border region runs out of computing resources with the baseline approach (as shown with dashed curves in Fig. \ref{fig:app_base_garr}(a)). The probability curves in Fig. \ref{fig:app_base_garr}(b) begin to re-converge at load 12000, when the border region with PESS also becomes full. Solid curves in Fig. \ref{fig:app_base_garr}(a) indicate that, between loads 1000 and 6000, when the blocking probability of the two experiments is comparable, \ac{pess} requires around 50\% less computing resources than the baseline to provision the security services.

\revised{
The results obtained with the Stanford network model are presented in Fig. \ref{fig:app_base_stanford}. In contrast to the GARR network, where busy links in sparsely connected areas cause rejected requests at low loads, in these experiments we see a non-zero blocking probability only when the border region of the Stanford network runs out of computing resources, i.e. at loads 4000 and 6000 for the baseline and \ac{pess}, respectively. \\

Similar to the random networks scenario, we can observe a higher number of active services and a lower end-to-end latency with \ac{pess} in both GARR and Stanford networks. The plots are omitted due to space constraints.
}

\subsection{Scalability evaluation}\label{sec:scalability}
\textbf{Methodology.} We evaluate the scalability of the \ac{pess} heuristic on Barab{\'a}si-Albert random topologies of between 10 and 1000 nodes. For each of these topologies, we simulate the processing of 1000 service requests and report the average execution time.

\textbf{Metrics.} Average execution time.

\textbf{Discussion.} In the first experiment (reported in the leftmost plot of Fig.~\ref{fig:scalability-plots}) we used $|EP2|=1$ for all the service requests and we varied the attachment parameter $m$, which determines the number of edges to attach from a new node to existing nodes when generating the random network. This influences the execution time of the shortest path algorithm. For instance, $m=1$ produces tree-like topologies with $|E|=|N|-1$. The general rule for computing the number of edges in Barab{\'a}si-Albert networks is $|E|=m\cdot|N|-m^2$.
As illustrated in Fig. \ref{fig:scalability-plots}, even for very large networks with 1000 nodes and 4975 edges ($m=5$ in the figure), on average, the PESS heuristic can provision a security service in around 200\,ms.

In the second experiment, we used a fixed value of $m=5$ (the worst case in the first experiment) and we varied the size of endpoint $EP2$, as the number of nodes in $EP2$ determines how long PESS takes to compute the initial solution. In the rightmost plot in Fig.~\ref{fig:scalability-plots}, the black solid curve is the reference measurement from the first experiment. As shown by the dashed curves in the plot, the average execution time increases linearly with the size of endpoint $|EP2|$, up to around 250\,ms in the worst case with $|N|=1000$, $|E|=4975$ and $|EP2|=500$.

As introduced in Section \ref{sec:introduction} and formulated in Section \ref{sec:model}, the \acp{vsnf} placement model and heuristic proposed in this work target NFV-enabled systems where security services are dynamically provisioned and updated based on users' applications and their security and \ac{qos} requirements. Such systems require efficient provisioning strategies to minimize the exposure of such applications to cyber attacks. With respect to these objectives, the experimental results from the PESS scalability evaluation are encouraging and clearly indicate the potential for practical implementation of the proposed application-aware approach in real-world scenarios. 

\section{Conclusions}\label{sec:conclusions}
In this paper, we have tackled the problem of the progressive provisioning of security services by means of \acp{vsnf}. The proposed approach, called \ac{pess}, takes into account  security and \ac{qos} requirements of user applications, while ensuring that computing and network resources are accurately utilised. We have discussed the rationale behind our design decisions and presented an \ac{ilp} formulation and a heuristic algorithm that solve the placement problem. Although we have focused our work on security services, the \ac{pess} approach is applicable to more complex scenarios, where heterogeneous network services provided by means of generic \acp{vnf} coexist. 

The evaluation results demonstrate the benefits of \ac{pess} for both users and operators, with savings in resource utilization and in end-to-end latency. We have also shown that the heuristic implementation of the proposed application-aware approach produces near-optimal solutions and scales well in large and dense networks, indicating the potential of \ac{pess} in real-world scenarios.

\bibliographystyle{IEEEtran} 
\bibliography{bibliography}

\begin{thebibliography}{10}
\providecommand{\url}[1]{#1}
\csname url@samestyle\endcsname
\providecommand{\newblock}{\relax}
\providecommand{\bibinfo}[2]{#2}
\providecommand{\BIBentrySTDinterwordspacing}{\spaceskip=0pt\relax}
\providecommand{\BIBentryALTinterwordstretchfactor}{4}
\providecommand{\BIBentryALTinterwordspacing}{\spaceskip=\fontdimen2\font plus
\BIBentryALTinterwordstretchfactor\fontdimen3\font minus
  \fontdimen4\font\relax}
\providecommand{\BIBforeignlanguage}[2]{{%
\expandafter\ifx\csname l@#1\endcsname\relax
\typeout{** WARNING: IEEEtran.bst: No hyphenation pattern has been}%
\typeout{** loaded for the language `#1'. Using the pattern for}%
\typeout{** the default language instead.}%
\else
\language=\csname l@#1\endcsname
\fi
#2}}
\providecommand{\BIBdecl}{\relax}
\BIBdecl

\bibitem{7243304}
{R. Mijumbi et al.}, ``{Network Function Virtualization: State-of-the-Art and
  Research Challenges},'' \emph{IEEE Communications Surveys Tutorials},
  vol.~18, no.~1, pp. 236--262, 2016.

\bibitem{5384976}
{R. Hill et al.}, ``{Measuring Latency for Video Surveillance Systems},'' in
  \emph{Proc. of DICTA}, 2009.

\bibitem{Claypool:2006:LPA:1167838.1167860}
{M. Claypool et al.}, ``{Latency and Player Actions in Online Games},''
  \emph{Comm. of the ACM}, vol.~49, no.~11, Nov. 2006.

\bibitem{Chen:2004:QRN:1234242.1234243}
{Y. Chen et al.}, ``{QoS Requirements of Network Applications on the
  Internet},'' \emph{Inf. Knowl. Syst. Manag.}, vol.~4, no.~1, pp. 55--76, Jan.
  2004.

\bibitem{Claypool:2010:LKP:1730836.1730863}
{M. Claypool et al.}, ``{Latency Can Kill: Precision and Deadline in Online
  Games},'' in \emph{Proc. of ACM MMSys}, 2010.

\bibitem{short-paper}
{R. Doriguzzi-Corin et al.}, ``{Application-Centric Provisioning of Virtual
  Security Network Functions},'' in \emph{Proc. of the Third IEEE International
  Workshop on Security in NFV-SDN (SN-2017)}, 2017.

\bibitem{7469866}
{F. Bari et al.}, ``{Orchestrating Virtualized Network Functions},'' \emph{IEEE
  TNSM}, vol.~13, no.~4, pp. 725--739, 2016.

\bibitem{6968961}
{S. Mehraghdam et al.}, ``{Specifying and Placing Chains of Virtual Network
  Functions},'' in \emph{Proc. of IEEE CloudNet}, 2014.

\bibitem{qos-driven}
{P. Vizarreta et al.}, ``{QoS-driven Function Placement Reducing Expenditures
  in NFV Deployments},'' in \emph{Proc. of ICC}, 2017.

\bibitem{8480442}
{M. M. Tajiki et al.}, ``{Joint Energy Efficient and QoS-aware Path Allocation
  and VNF Placement for Service Function Chaining},'' \emph{IEEE TNSM}, 2018.

\bibitem{Park:2017:DDP:3040992.3041005}
{Y.Parket al.}, ``{Dynamic Defense Provision via Network Functions
  Virtualization},'' in \emph{Proc. of ACM SDN-NFV Sec.}, 2017.

\bibitem{7899497}
{T. V. Phan et al.}, ``{Optimizing resource allocation for elastic security
  VNFs in the SDNFV-enabled cloud computing},'' in \emph{Proc. of ICOIN}, 2017.

\bibitem{8530989}
{S. Demirci et al.}, ``{Optimal Placement of Virtual Security Functions to
  Minimize Energy Consumption},'' in \emph{Proc. of the IEEE ISNCC}, 2018.

\bibitem{8254344}
{Q. Xu et al.}, ``{Low Latency Security Function Chain Embedding Across
  Multiple Domains},'' \emph{IEEE Access}, 2018.

\bibitem{7592416}
{A. Shameli Sendiet al.}, ``{Efficient Provisioning of Security Service
  Function Chaining Using Network Security Defense Patterns},'' \emph{IEEE
  Transactions on Services Computing}, 2017.

\bibitem{8466784}
{Y. Liu et al.}, ``{A Dynamic Composition Mechanism of Security Service
  Chaining Oriented to SDN/NFV-Enabled Networks},'' \emph{IEEE Access}, vol.~6,
  pp. 53\,918--53\,929, 2018.

\bibitem{huang}
{D. Huang et al.}, \emph{Software-Defined Networking and Security: From Theory
  to Practice}.\hskip 1em plus 0.5em minus 0.4em\relax CRC Press, 2018.

\bibitem{openflow}
{N. McKeown et al.}, ``{OpenFlow: enabling innovation in campus networks},''
  \emph{{ACM SIGCOMM Computer Communication Review}}, vol.~32, no.~2, pp.
  69--74, April 2008.

\bibitem{Gianvecchio:2011:HBI:2109150.2109174}
{S. Gianvecchio et al.}, ``{Humans and Bots in Internet Chat: Measurement,
  Analysis, and Automated Classification},'' \emph{IEEE/ACM Trans. Netw.},
  vol.~19, no.~5, pp. 1557--1571, Oct. 2011.

\bibitem{Yan:2005:SCC:1103599.1103606}
{J. Yan et al.}, ``{A Systematic Classification of Cheating in Online Games},''
  in \emph{Proc. of ACM SIGCOMM NetGames}, 2005.

\bibitem{Sherry:2015:BDP:2785956.2787502}
{J. Sherry et al.}, ``{BlindBox: Deep Packet Inspection over Encrypted
  Traffic},'' in \emph{Proc. of ACM SIGCOMM}, 2015.

\bibitem{Canard:2017:BMP:3052973.3053013}
{S. Canard et al.}, ``{BlindIDS: Market-Compliant and Privacy-Friendly
  Intrusion Detection System over Encrypted Traffic},'' in \emph{Proc. of ACM
  ASIA CCS}, 2017.

\bibitem{nfvipop}
{ETSI}, ``{Network Functions Virtualisation (NFV); Terminology for Main
  Concepts in NFV},''
  \url{https://www.etsi.org/deliver/etsi_gs/NFV/001_099/003/01.02.01_60/gs_NFV003v010201p.pdf},
  [Accessed: 31-May-2019].

\bibitem{mirai}
{M. Antonakakis et al.}, ``{Understanding the Mirai Botnet},'' in \emph{USENIX
  Security Symposium}, 2017.

\bibitem{Sherali2013}
{H.D. Sherali et al.}, ``Reformulation--linearization techniques for discrete
  optimization problems,'' in \emph{Handbook of Combinatorial Optimization},
  P.~M. Pardalos, D.-Z. Du, and R.~L. Graham, Eds.\hskip 1em plus 0.5em minus
  0.4em\relax Springer US, 2013, pp. 2849--2896.

\bibitem{8704949}
{M. Savi et al.}, ``{Impact of Processing-Resource Sharing on the Placement of
  Chained Virtual Network Functions},'' \emph{IEEE Transactions on Cloud
  Computing}, 2019.

\bibitem{GAO2018108}
{Meihui Gao et al.}, ``Optimal orchestration of virtual network functions,''
  \emph{Computer Networks}, vol. 142, pp. 108 -- 127, 2018.

\bibitem{oljira-qos-aware}
{D.B. Oljira et al.}, ``{A Model for QoS-Aware VNF Placement and
  Provisioning},'' in \emph{Proc. of IEEE NFV-SDN}, Nov 2017.

\bibitem{7842188}
{F. Ben Jemaa et al.}, ``{QoS-Aware VNF Placement Optimization in Edge-Central
  Carrier Cloud Architecture},'' in \emph{Proc. of IEEE GLOBECOM}, 2016.

\bibitem{silo}
{K. Jang et al.}, ``{Silo: Predictable Message Latency in the Cloud},'' in
  \emph{Proc. of the 2015 ACM Conference on Special Interest Group on Data
  Communication}, 2015.

\bibitem{gurobi_sos}
{Gurobi Optimization}, ``{Constraints},''
  \url{http://www.gurobi.com/documentation/7.5/refman/constraints.html},
  [Accessed: 31-May-2019].

\bibitem{gurobi}
``{Gurobi Optimizer},'' \url{http://www.gurobi.com}, [Accessed: 31-May-2019].

\bibitem{gurobi_api}
{Gurobi Optimization}, ``{Python API},''
  \url{http://www.gurobi.com/documentation/7.5/refman/py_python_api_overview.html},
  [Accessed: 31-May-2019].

\bibitem{barabasi}
\BIBentryALTinterwordspacing
A.-L. Barab{\'a}si and R.~Albert, ``{Emergence of Scaling in Random
  Networks},'' \emph{Science}, vol. 286, no. 5439, pp. 509--512, 1999.
  [Online]. Available: \url{http://science.sciencemag.org/content/286/5439/509}
\BIBentrySTDinterwordspacing

\bibitem{headerspace}
{P. Kazemian et al.}, ``{Header Space Analysis: Static Checking for
  Networks},'' in \emph{Proc. of the 9th USENIX Conference on Networked Systems
  Design and Implementation}, 2012.

\bibitem{garr-whitepaper}
{GARR}, ``{Considering the Next Generation of GARR Network},''
  \url{https://www.garr.it/it/documenti/3474-garr-white-paper-maggio-2017},
  2017, [Accessed: 31-May-2019].

\bibitem{garr-topology}
GARR, ``{GARR Network Map},''
  \url{https://www.garr.it/images/2018-01_mappaGARR_Net.png}, 2018, [Accessed:
  31-May-2019].

\bibitem{garr-backbone}
{GARR}, ``{GARR Backbone},''
  \url{https://www.garr.it/en/infrastructures/network-infrastructure/backbones},
  [Accessed: 31-May-2019].

\bibitem{SHAH2018157}
{S.A.R. Shah et al.}, ``{Performance comparison of intrusion detection systems
  and application of machine learning to Snort system},'' \emph{Future
  Generation Computer Systems}, vol.~80, no. Supplement C, pp. 157 -- 170,
  2018.

\bibitem{7973470}
{D. Lackovi\'c et al.}, ``{Performance analysis of virtualized VPN
  endpoints},'' in \emph{Proc. of 40th MIPRO}, May 2017.

\bibitem{fortigate}
{Fortinet}, ``{Fortigate Virtual Applicances},''
  \url{https://www.fortinet.com/content/dam/fortinet/assets/data-sheets/FortiGate_VM.pdf},
  2018, [Accessed: 31-May-2019].

\bibitem{cisco_asav}
{Cisco Systems}, ``{Adaptive Security Virtual Appliance (ASAv)},''
  \url{https://www.cisco.com/c/en/us/products/collateral/security/adaptive-security-virtual-appliance-asav/datasheet-c78-733399.pdf},
  2018, [Accessed: 31-May-2019].

\bibitem{juniper_vsrx}
{Juniper Networks}, ``{vSRX Virtual Firewall},''
  \url{https://www.juniper.net/assets/us/en/local/pdf/datasheets/1000489-en.pdf},
  2018, [Accessed: 31-May-2019].

\end{thebibliography}

\end{document}